\pdfoutput=1
\documentclass[journal,onecolumn,12pt]{IEEEtran}
\usepackage[noadjust]{cite}
\usepackage{multirow}
\usepackage{makecell}
\usepackage{graphicx}
\usepackage{booktabs}
\usepackage{color}
\usepackage{rotating}
\usepackage{algorithm}
\usepackage{algorithmicx}
\usepackage{algpseudocode}
\usepackage{amsthm,amsmath,amssymb,lipsum}
\usepackage{mathrsfs}
\usepackage{multicol}
\usepackage{lineno}
\usepackage{epstopdf}

\begin{document}

\title{Control-Oriented Modelling and Adaptive Parameter Estimation for Hybrid Wind-Wave Energy Systems}

\author{Yingbo Huang, Bozhong Yuan, Haoran He, Jing Na, Yu Feng, Guang Li, Jing Zhao, Pak Kin Wong and Lin Cui
\thanks{
  Yingbo Huang, Bozhong Yuan, Haoran He and Jing Na are with Faculty of Mechanical and Electrical Engineering, Yunnan Key Laboratory of Intelligent Control and Application, Kunming University of Science and Technology, Kunming, 650500, Yunnan, China. (e-mail: Yingbo\_Huang@126.com). Yu Feng is with Chongqing Haizhuang Windpower Engineering and Research Co., LTD, 400021, Chongqing, China. Guang Li is with Faculty of Mechanical and Electrical Engineering, Kunming University of Science and Technology, Kunming, 650500, Yunnan, China and School of Engineering, The University of Manchester, Manchester, M13 9PL, UK. Jing Zhao is with Department of Electromechanical Engineering, University of Macau, Taipa, 999078, Macau, China and Zhuhai UM Research Institute, Zhuhai, 519031, Guangdong, China. Pak Kin Wang is with Department of Electromechanical Engineering, University of Macau, Taipa, 999078, Macau, China and Zhuhai UM Research Institute, Zhuhai, 519031, Guangdong, China. Lin Cui is with Jiangsu Marine Technology Innovation Center, Nantong, 226100, Jiangsu, China.
  }
}

\maketitle

\begin{abstract}
  Hybrid wind-wave energy system, integrating floating offshore wind turbine (FOWT) and wave energy converters (WECs), has received much attention in recent years due to its potential benefit in increasing the power harvest density and reducing the levelized cost of electricity (LCOE). Apart from the design complexities of the hybrid wind-wave energy systems, their energy conversion efficiency, power output smoothness and their safe operations introduce new challenges for their control system designs. Recent studies show that advanced model-based
  control strategies have the great potential to significantly improve their overall control performance. However the performance of these advanced control strategies rely on the computationally efficient control-oriented models with sufficient fidelity, which are normally difficult to derive due to the complexity of the hydro-, aero-dynamic effects and the couplings.In
  most available results, the hybrid wind-wave energy system models are established by using the Boundary Element Method (BEM), devoting to understanding the hydrodynamic responses and performance analysis. However, such models are complex and involved relatively heavy
  computational burden, which cannot be directly used for the advanced model-based control methods that are essential for improving power capture efficiency from implementing in practice. To overcome this issue, this paper proposes a control-oriented model of the hybrid windwave
  energy system with six degrees of freedom (DOFs). First, the Newton's second law and fluid mechanics are employed to characterize the motion behaviour of the hybrid wind-wave energy system with the coupled aero-hydro-mooring dynamics. Then, a novel adaptive parameter estimation algorithm with simple low-pass filter approach is developed to estimate the system unknown coefficients. Different from the conventional parameter estimation methods, such as gradient descent method and recursive least-squares (RLS) method, the estimated parameters can be driven to their true values with guaranteed convergence. Finally, numerical analysis using the AQWA and MATLAB are applied to validate the fidelity of the control-oriented model under different wind and wave conditions. The results indicate that the control-oriented model predicts
  the motion response accurately in comparison to the BEM-based model. Overall, the results pave the way for designing advanced hybrid wind-wave energy system control method.
\end{abstract}


\section{Introduction}\label{sec1}
Given the fact that the traditional fossil energy sources are decreasing, exploring renewable energy emerges an inevitable trend in all over the world. Considering the various renewable energy resources, offshore wind energy has been regarded as one of the most promising renewable energies. This is because the onshore wind energy is an early-developed renewable energy, some key technologies of which, for instance rotor technology, is quite mature and can be employed for offshore wind energy capture directly. It has been well-accepted that best offshore wind energy resource is located on deep sea (more than 40 \text{m}) with power intensity of 60-70 kW/m. Therefore, the floating offshore wind turbine (FOWT) is developed and considered as a viable and economical alternative to bottom-mounted wind turbine for harnessing offshore wind energy \cite{shi2023real} . Although the deep sea areas have stable and abundant wind resources, some critical issues, such as high wind speed and high wave, may make the turbines move violently, leading to large external loads on the structure and drastically reducing the service life of FOWT. As discussed in literature \cite{Suzuki2007Load}, the FOWT pitching with an amplitude of 5$^{\circ}$ will impose a 50$\%$ increase requirement on blade sectional modulus to mitigate the fatigue failure risk.

Given the fact that ocean wave and offshore wind show a high positive correlation in spatial and temporal
distribution \cite{de2024assessment}, the idea of combination of FOWT and wave energy converter (WEC) was proposed and the hybrid wind-wave energy system is presented \cite{perez2015review}. Specifically, the hybrid system indicates that FOWT and WEC share the same floating platform and the WEC is arranged around the floating platform and acted as an extra spring-damper system of FOWT. In this manner, the relative motion between FOWT and WEC can be regulated in terms of passive or active control methods. Among the several types of floating platform (i.e., spar, barge, tension leg platform and semi-submersible platform) \cite{zhang2024computational}, semi-submersible platform attracts great attentions due to its flexibility in deployment for various water depth. It should be noted that the largest FOWT named WindFloat (25 MW) adopts the semi-submersible platform. Although the hybrid wind-wave energy system holds advantages from both FOWT and WEC, few commercial-scale applications have been reported. The reason behind this fact comes from many factors, but one primary reason is that the WEC technologies are still at nascent stage. In 2015, Nielsen et al. \cite{nielsen2015partnership} evaluated the performance of various WECs and the test results show that the Wave Star has the highest Technology Readiness Level (TRL) and lowest levelized cost of electricity (LCOE). In this sense, the combination of semi-submersible floating wind turbine and Wave Star presents a great potential for development \cite{Peng2023Optimization}.

However, the design of hybrid energy system poses significant challenges due to their intricate structure, extensive coupling effects and highly complex environmental loads, etc. In the assessment of the economic benefits and development cycle, tank testing is an essential step. Si et al. \cite{zhang2022coupled} conducted experiments on the hybrid energy system with 1:50 scale ratio in wave basin, the obtained results confirm the feasibility of the hybrid energy system. In tank testing, satisfying the Froude scaling law and Reynolds similarity law simultaneously is not an easy task, resulting in the disparity issue between the scale physical model and actual model. However, the numerical model proposed for the actual model is not affected by the scaling ratio. Therefore, establishing an accurate mathematical model is quite important to better understand the dynamic responses and performance analysis of the hybrid wind-wave energy system under different marine environments. The widely used numerical simulation software is the OpenFAST developed by the National Renewable Energy Laboratory (NREL), by which both the wind turbine load and motion responses of floating platform can be calculated and the corresponding system characterization can be analysed \cite{Jonkman2005FAST}. Si et al. \cite{si2021influence} used AQWA to model the DeepCwind-Wavestar-Combined (DWC) system and compared the system dynamics with OpenFAST to illustrate the fidelity of the established model in AQWA. Zhu \cite{Zhu2023Optimal} adopted SimMechanics toolbox in Matlab to model the hybrid wind-wave energy system, by which the bodies, joints and constraints can be replaced by blocks, and the motion responses of the whole system can be obtained by connecting ports of the blocks. Yang et al. \cite{YANG2020Development} developed a toolkit named F2A (OpenFAST to AQWA) to realize the data transmission between the two software via dynamic link libraries (DLL), by which the corresponding post-processed on this basis can be conducted. Micallef et al. \cite{micallef2017dynamic} simulated the hybrid system with waves using the Boundary Element Method (BEM) and relied on this to develop the numerical model. Although the above numerical simulations yield satisfactory results, the adopted mathematical models of the hybrid wind-wave energy system are generally established via the BEM. It is true that the BEM-based model can describe the system dynamics comprehensively due to its high fidelity, but it is more suitable for hydrodynamic analysis instead of the controller design as the complex model and relatively high computational burden. Liu et al. \cite{liu2024optimization} employed the Artificial Neural Networks (ANNs) to approximate the nonlinear dynamics encountered into the hybrid energy system. However, one of the well-known drawbacks of the ANNs is that it requires a substantial amount of data and lacks interpretability. Peter et al. \cite{stansby2024wind} proposed a time domain linear diffraction–radiation model, which significantly improves calculation speed compared to previous commercial software. However, this model primarily focuses on the study of structural optimization and layout improvement.

Based on the above discussions, this paper proposes a control-oriented model of the hybrid wind-wave energy system based on the Newton's second law and fluid mechanics. Unlike the existing results that only consider two dimensions, the established model has six degrees of freedom (DOFs) by taking the coupled aero-hydro-mooring dynamics into account. In this sense, the dynamic responses of the hybrid wind-wave energy system can be simulated by the proposed control-oriented model more comprehensively. To obtain a high-fidelity control-oriented model of the hybrid energy system, a new adaptive parameter estimation algorithm is suggested to online estimate the uncertain parameters of the control-oriented model via designing a set of low-pass filer operations and auxiliary matrices. One salient feature of this parameter estimation method is that the potential bursting phenomenon encountered into the gradient descent method and recursive least-squares (RLS) can be remedied, such that the parameter convergence can be guaranteed, i.e., the estimated parameters can be driven to their true values. Numerical analyses are conducted based on the AQWA-Matlab framework, where the motion responses of the BEM-based model and control-oriented model are compared under various wind and wave conditions to demonstrate the fidelity of the proposed control-oriented model. 

The structure layout of this paper is organized as follows. Section 2 describes the concept of the designed hybrid wind-wave energy system and its key parameters are also provided. The mathematical development and control-oriented model of the hybrid wind-wave energy system are given in Section 3. Section 4 provides the details of the adaptive parameter estimation algorithm. Section 5 introduces the AQWA-Matlab framework and the comparative numerical results and the corresponding explanations are given in Section 5. Conclusions and future works are claimed in Section 6.

\section{Hybrid wind-wave energy system description}
In this section, the proposed hybrid wind-wave energy system will be described and key parameters of the hybrid system will be provided as well.

\begin{itemize}
\item\textbf{\textit{DeepCwind:}} DeepCwind, as shown in Fig. 1(a), is a semi-submersible type FOWT developed by the NREL, which consists of multiple columns. One primary column (generally center column) is connected to the tower and three offset columns are connected to the primary column in terms of  multiple trusses \cite{imran2017optimal}. The NREL 5 MW baseline wind turbine is mounted on the tower located on the center column. The mooring system of Deepcwind has three steel catenary lines, which are placed at 120$^{\circ}$ relative to each other. Owing to this structure, the DeepCwind can effectively avoid excessive heave motion of the floating platform. Generally, the generator torque control and pitch blade control are implemented on wind turbines to maximize the power tracking and power regulation when operating conditions are not rated. The detailed parameters of DeepCwind are given Table~\ref{tab1}.

\begin{table}[htb]
  \centering
  \caption{Parameters of DeepCwind.}\label{tab1}
\tabcolsep=1.3cm \renewcommand\arraystretch{1.1}
  \begin{tabular}{@{}llll@{}}
 \toprule
    ~ & \textbf{Parameters}  & \textbf{Value} & \textbf{Unit}  \\
    \midrule
    Wind turbine& Rated power   & 5 & MW  \\ ~& Center of mass (CM) & (-0.2, 0.0, 70.0) & m  \\
~ & Rotor diameter & 126 & m  \\
~ & Hub height & 90 & m \\
~ & Rotor mass & 110000 & kg  \\
~ & Nacelle mass & 240000& kg \\
~ & Tower mass & 347460 & kg \\ 
\midrule
Platform & CM & (0.0, 0.0, -13.46) & m \\  
 ~ & Mass & $1.1473\times10^7$ & kg \\ 
~ & Roll/ pitch inertia about CM & $6.827\times10^9$ & kg m$^2$ \\  
~ & Yaw inertia about CM & $1.226\times10^{10}$ & kg m$^2$ \\ 
\midrule
Mooring & Depth to anchors below SWL & 200 & m \\
~ & Depth to fairleads below SWL & 14 & m \\ 
 ~ & Unstretched mooring line length & 835.5 & m \\ 
 ~ & Mooring line diameter & 0.0766 & m \\ 
~ & Equivalent mass density & 113.35 & kg/m \\ 
 ~ & Equivalent extensional stiffness & $7.536\times10^8$ & N \\ 
 \bottomrule
\end{tabular}
\end{table}

\end{itemize}

\begin{itemize}
\item\textbf{\textit{Wavestar:}} Wavestar developed by Aalborg University, as shown in Fig. 1(b), has been widely-accepted as one of the leading commercial devices for wave power generation. The device consists of a hemispherical buoy with a ballast tank inside, by which the natural frequency of the buoy can be adjusted in terms of injecting water. The buoy is fixed on the horizontal axis through the connecting arm. Under operating condition, the buoy is driven by the ocean wave and then the corresponding reciprocating rotating motion drives the power take-off (PTO) to generate electricity. When suffering from the hostile marine environment, the equipped hydraulic system lifts the buoy out of the water to ensure equipment safety. Some key parameters of the Wavestar can be found in Table~\ref{tab2}.

\begin{table}[htb]
  \centering
  \caption{Parameters of WaveStar.}\label{tab2}
\tabcolsep=1.4cm \renewcommand\arraystretch{1.1}
  \begin{tabular}{@{} lll@{} }
\toprule
\textbf{WEC Parameters}  & \textbf{Value} & \textbf{Unit}   \\
\midrule
 Diameter at the SWL & 9.8 & m \\
Elevation of WECs above SWL & 3.2 & m \\ 
CM of WEC1 & (-28.78, 49.85, -1.45) & m \\ 
CM of WEC2 & (-28.78, -49.85, -1.45) & m \\ 
CM of WEC3 & (57.67, 0.00, -1.45) & m \\ 
Mass & 18827 & kg \\ 
 Roll/ pitch inertia about CM & $1.029\times10^6$ & kg m$^2$ \\ 
Yaw inertia about CM & $1.704\times10^6$ & kg m$^2$ \\
\bottomrule
\end{tabular}
\end{table}

\end{itemize}

This paper integrates the Wavestar \cite{RANSLEY201749} on the semi-submersible type FOWT 5MW DeepCwind  \cite{Robertson2014Definition} to characterize the hybrid wind-wave energy system, the conceptual design of which is given in Fig. 1(c). Three point-absorber type WECs are arrayed around the floating platform at 120$^\circ$ to each other with the platform as the center of the circle. Each of the three point-absorbers is connected to the three columns of the floating platform (as depicted in Fig. 1(c) ) by connecting arms. Since the connecting arms are connected to the columns by hinged structures, there is only relative rotation to each other. One should be pointed out that the increased number of WECs may contribute to yielding the energy capture efficiency improvement of the hybrid energy systems, but for the ease of the subsequently analysis, only three WECs are integrated on the floating platform. 
\begin{figure*}[!htbp]
	\centering
	\includegraphics[width=13cm, height=11cm]{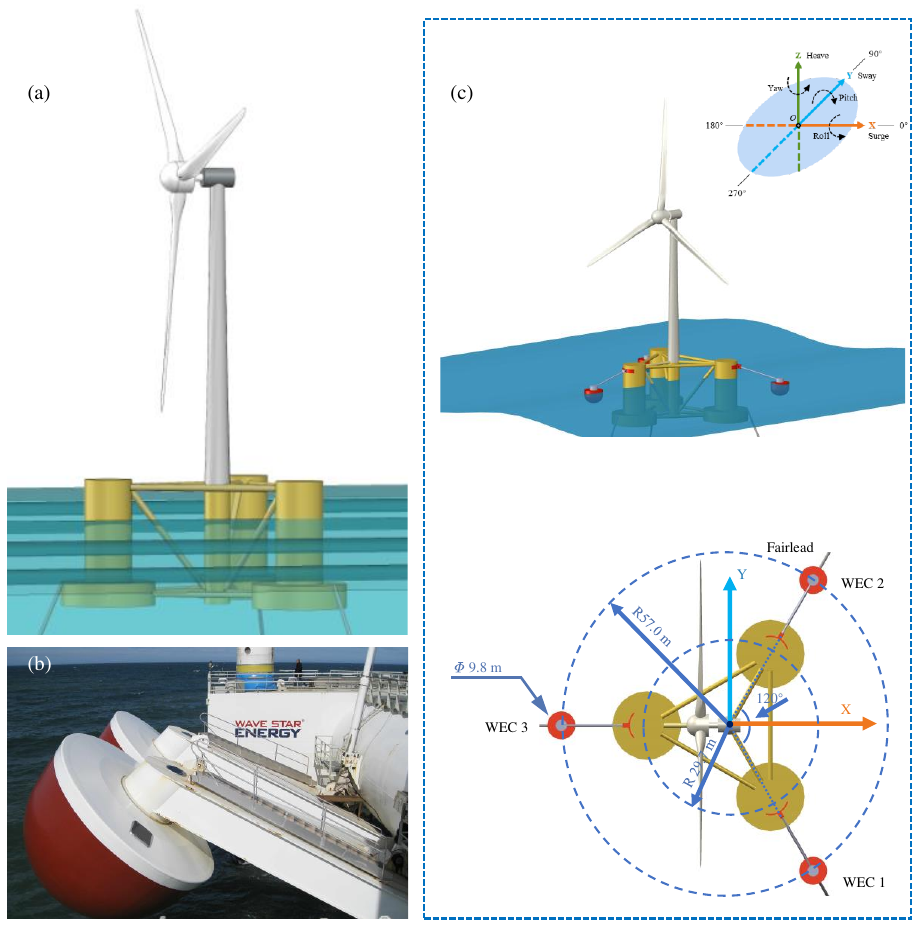}
	\caption{\fontfamily{ptm}\selectfont (a) DeepCwind, (b) Wavestar and (c) Hybrid wind-wave energy system.}
	\label{fig1}
\end{figure*}

As discussed above, in the hybrid wind-wave energy system, the employed WEC device can be regarded as the spring-damper system to maintain the stability of the floating platform and thereby contribute to improving energy capture efficiency. Apart from the proper mechanical design, advanced control method has also been recognized as a powerful tool to improve the floating platform stability, which has been widely accepted in ocean research community. In order to implement the control method successfully and achieve satisfactory control result, establishing a higher fidelity mathematical model of the hybrid wind-wave energy system is of vita importance. Generally, the model of hybrid wind-wave energy system can be classified in two types, i.e., BEM-based model and control-oriented model. The main merit of the BEM-based model lies in that it can characterize the dynamic behaviour more comprehensively, which benefits to conducting hydrodynamic responses analysis. Compared with the BEM-based model, the control-oriented model uses the existing mathematical methods of physics to establish a series of governing differential equation to describe the system responses. In this manner, the control-oriented model has relatively low model order and complexity. However, due to the reduced model order and complexity, the unmodeled dynamics and/or parametric model uncertainties emerge, especially for the involved parameter uncertainties that may prevent the control-oriented model from generating desired system responses. In this respect, this paper devotes to establishing an accurate control-oriented model of the hybrid wind-wave energy system with a novel adaptive parameter estimation algorithm.

\section{Hybrid wind-wave energy system model}

In this section, a detailed dynamic model of the hybrid wind-wave energy system is presented, where the interactions between the FOWT and the WECs are also modelled along with detailed descriptions about how the external forces are calculated. Then, the control-oriented model of the hybrid wind-wave energy system is derived by choosing state variables appropriately. 

\subsection{Dynamic model}
The strong interaction between the platform and the WECs imposes significant impact on each other's motion response. Therefore, the coupled aero-hydro-mooring dynamics between the platform and the WECs need to be considered when modelling the hybrid energy system. Note that both the platform and the WECs are treated as rigid bodies, such that the whole hybrid energy system has six DOFs. Since the platform and the WECs are mechanically coupled together, the platform and the three WECs are treated as a single rigid body in the surge, sway and yaw modes \cite{TIAN2023113824}. Meanwhile, the motion response of the WECs are significantly influenced by the pitch, roll, and heave modes of the platform, hence the hybrid energy system cannot be considered as a single rigid body in the pitch, roll and heave modes. Note that the reference coordinate defined in this paper is autonomous and generalized. The letters $x, y$ and $z$ represent the three orthogonal axes of the reference coordinate system. The $z$-axis is oriented vertically upwards along the center line of the tower, while the $x$-axis aligns with the mean wind and wave direction, the orientation of the $y$-axis follows the right-hand rule. The $xy$-plane is defined as still water level (SWL).

Due to the symmetrical configuration of the hybrid wind-wave energy system, only the motion equations of the platform in the surge, pitch and heave modes are studied in this paper. In subsequent sections, the subscripts $i=1,2,3$ represent the FWOT surge, pitch and heave modes respectively, and $i=4,5,6$ represent the pitch modes of each of the three WECs. Based on the Lagrange's equations and the Newton's second law, the dynamic equation of the hybrid wind-wave energy system is described as follows:
\begin{equation}
  \label{eq1}
M_i\ddot{q}(t)+B_i\dot{q}\left( t \right) +R_iq(t)+C_i=F_{i}^{exe}(t)
\end{equation}
where $\ddot{q}(t), \dot{q}(t), q(t)\in \mathbb{R}^{6\times 1}$ represent the vectors of acceleration, velocity and displacement respectively. 
$M_i=\mathrm{diag}\mathrm{(}M_1,M_2,M_3,M_4,M_5,M_6)\in \mathbb{R}^{6\times 6}$ represents mass matrix, where $M_{1}=M_{3}$ denotes the platform mass. $M_{2}$ represents the pitch inertia of platform about CM. $M_{4}, M_{5}, M_{6}$ are the three WEC's pitch inertia relative to their own CM respectively. Since the presence of gravity in the heave mode as well as the pre-tension of the mooring system, a constant vector $C_i=\left[ 0,0,C_3,0,0,0 \right] ^T\in \mathbb{R}^{6\times 1}$ is introduced. $F_{i}^{exe}(t)\!=\!\left[ F_{1}^{exe}(t),F_{2}^{exe}(t), F_{3}^{exe}(t), F_{4}^{exe}(t),F_{5}^{exe}(t),F_{6}^{exe}(t) \right] ^T$ $\in \mathbb{R}^{6\times 1}$ denotes the lumped external force. $B_i\in \mathbb{R} ^{6\times 6}$ and $R_i\in \mathbb{R} ^{6\times 6}$ denote the damping coefficient matrix and stiffness coefficient matrix of the hybrid energy system respectively. The detailed mathematical developments of $F_{i}^{exe}(t), B_i$ and $R_i$ are given as follows:
\begin{equation}\label{eq2}
F_{i}^{exe}(t)=\begin{bmatrix}F_{1}^{exe}(t)\\F_{2}^{exe}(t)\\F_{3}^{exe}(t)\\
F_{4}^{exe}(t)\\F_{5}^{exe}(t)\\F_{6}^{exe}(t)\end{bmatrix}
=
\begin{bmatrix}F_{1}^{aero}(t)+F_{1}^{rad}(t)+F_{1}^{exc}(t)+F_{1}^{hydro}(t)+F_{1}^{moor}(t)\\
F_{2}^{aero}(t)+F_{2}^{rad}(t)+F_{2}^{exc}(t)+F_{2}^{hydro}(t)+F_{2}^{moor}(t)+F_{2}^{PTO}(t)\\
F_{3}^{aero}(t)+F_{3}^{rad}(t)+F_{3}^{exc}(t)+F_{3}^{hydro}(t)+F_{3}^{moor}(t)+F_{3}^{PTO}(t)+F_{3}^{gra}\\
F_{4}^{rad}(t)+F_{4}^{exc}(t)+F_{4}^{hydro}(t)+F_{4}^{PTO}(t)\\F_{5}^{rad}(t)+
F_{5}^{exc}(t)+F_{5}^{hydro}(t)+F_{5}^{PTO}(t)\\F_{6}^{rad}(t)+F_{6}^{exc}(t)+
F_{6}^{hydro}(t)+F_{6}^{PTO}(t)\end{bmatrix}
\end{equation}

\begin{equation}  \label{eq3}
B_i=\begin{bmatrix}B_{11}&0&0&0&0&0\\B_{21}&B_{22}&0&0&0&0\\0&0&B_{33}&0&0&0\\
B_{41}&B_{42}&0&B_{44}&0&0\\B_{51}&B_{52}&0&0&B_{55}&0\\
B_{61}&B_{62}&0&0&0&B_{66}\end{bmatrix},
R_i=\begin{bmatrix}R_{11}&0&0&0&0&0\\R_{21}&R_{22}&0&0&0&0\\0&0&R_{33}&0&0&0\\
R_{41}&R_{42}&0&R_{44}&0&0\\R_{51}&R_{52}&0&0&R_{55}&0\\
R_{61}&R_{62}&0&0&0&R_{66}\end{bmatrix}
\end{equation}
where $B_{ii}$ with $i=1\cdots6$ represents the damping coefficient of the platform and three WECs. $B_{21}$ represents the damping coefficient caused by the coupling effect between the platform surge and pitch. $B_{i1}, B_{i2}, i=4, 5, 6$ denote the damping coefficient stemmed from the coupling effect between platform and WECs respectively. The detailed explanations of the elements involved in $R_i$ are similar to  $B_i$.

As explained above, the hybrid energy system cannot be considered as a single rigid body in the pitch, roll, and heave modes, hence the forces on the different modes are not the same, and thereby $F_{i}^{exe}(t)$ can be further written as Eq. \eqref{eq2}. In \eqref{eq2}, $F_{i}^{aero}(t)$ denotes the the aerodynamic load caused by wind action that is affected by a lot of factors such as wind speed, wind direction and platform motion. Blade element momentum theory is widely used in wind turbine design, which divides the blade into multiple sections along the length and then calculates the aerodynamic loads for each unit. This is a complicated process to predict the aerodynamic load of three-dimensional marine structures accurately from the perspective of theoretical analysis. In order to simplify the process, we use AQWA to calculate the aerodynamic loads by wind drag coefficient and relative wind speed at the target height directly. The specific calculation method will not be repeated here, please refer to ANSYS for more details.

Furthermore, $F_{i}^{rad}$ is the radiation force due to the motion of each component of the system, which can be calculated as follows \cite{Cummins1962The}:
\begin{equation}
  \label{eq4}
F_i^{rad}(t)=-A_{\infty,ik}\ddot{x}_k(t)-\int_{0}^tK_{ik}(t-\tau)\dot{x}_k(\tau)d\tau 
\end{equation}
with ${A}_{\infty,ik}$ being the infinite-frequency added mass and the corresponding subscript $ik$ ($k$=1,2,3) denotes the coupling between the $i$th unit and $k$th unit. The convolution term $\int_0^{t}K_{ik}(t-\tau)\dot{x}_k(\tau)d\tau$ called fluid-memory model and $K_{ik}(t)$ represents the retardation impulse response function, which can be approximated by the following state-space model:
\begin{equation}
  \label{eq5}
\begin{cases}\dot{x}_{r}(t)=
A_{r}x_{r}(t)+B_{r}\dot{x}_k(t)\\\int_{0}^{t}K(t-\tau)\dot{x}_k(\tau)d\tau\approx C_{r}x_{r}(t)\end{cases}
\end{equation}
where $x_r\in\mathbb{R}^{n\times 1}$ denotes the state vector, the system matrix $A_r\in\mathbb{R}^{n\times n}$, input matrix $B_r\in\mathbb{R}^{n\times 1}$ and output matrix $C_r\in\mathbb{R}^{1\times n}$ can be obtained by system identification \cite{Prez2009IdentificationOD}.

$F_i^{exc}$ represents the wave excitation force that can be calculated by summing two force components (i.e., Froude-Krylov force and Diffraction force), which is given as:
\begin{equation}
  \label{eq6}
\begin{aligned}F_{i}^{exc}&=F_{FK}+F_{D}=\alpha\left|f(\omega)\right|e^{-j(\varpi t+\angle f(\omega))}\end{aligned}
\end{equation}
where $\alpha$ and $\varpi$ are the amplitude and frequency of incident waves. $|f(\omega)|$ and $\angle f(\omega)$ represent the amplitude and phase of the Froude-Krylov coefficient and the excitation load with respect to the incident wave, respectively. 

$F_{i}^{hydro}(t)$ denotes the fluid load acted on the object when the object is placed in still water, which is related to the fluid density, surface area and system states, etc. The mathematical development of $F_{i}^{hydro}(t)$ can be written as:
\begin{equation}
  \label{eq7}
F_{i}^{hydro}=\rho gV\delta -C_{ij}^{hydro}q_i
\end{equation}
where $\rho $ is the fluid density. $g$ is the gravitational acceleration. $V$ represents the volume of fluid displaced by the platform and $\delta $ is the identity matrix. $C_{ij}^{hydro}$ is the $(i,j)$ component of the linear hydrostatic-restoring matrix, which can be obtained from AQWA directly. The subscripts $i$ and $j$ represent six DOFs ranging from 1 to 6.

$F_{i}^{moor}(t)$ represents the force imposed on the platform by the mooring system, while the three WECs are attached to the platform and share a common mooring system with the platform. Note that the external force $F_{i}^{exe}(t), i=4,5,6$ does not include $F_{i}^{moor}(t)$. To reduce the model complexity of the mooring system, taking into account the speed of the hybrid energy system does not vary significantly, the dynamic equation of the mooring system can be linearized as follows \cite{Robertson2014Definition}:
\begin{equation}
  \label{eq8}
F_i^{moor}=\begin{cases}-R_mq_i-B_m\dot{q}_i,&i=
1,2\\-R_mq_i-B_m\dot{q}_i+c,&i=3\end{cases}
\end{equation}
where $R_m$ and $B_m$ are the stiffness coefficient and damping coefficient of mooring system, respectively. Note that when simulating pre-tension of the mooring system on the heave mode, a constant $c$ needs to be added in the case when $i$ = 3.

As shown in Fig. \ref{fig1}, three WECs are connected to the platform by revolute joint, so the $F_i^{PTO}$ can be expressed as \cite{da2022dynamics}:
\begin{equation}
  \label{eq9}
F_i^{PTO}=-\left.K_{PTO}q_{ri}-B_{PTO}\dot{q}_{ri}-u\left(q_r,\dot{q}_r\right)\right.
\end{equation}
where $q_r$ and $\dot{q}_r$ represent relative motion and velocity between platform and WECs. $u$ denotes the control input of PTO actuator. $K_{PTO}$ and $B_{PTO}$ are PTO stiffness coefficient and damping coefficient, respectively. $q_r$ can be calculated as follows:
\begin{equation}
  \label{eq10}
q_{ri}=q_{si}-q_{wi}
\end{equation}
where $q_{si}$ and $q_{wi}$ represent the rotation angle of the platform and each WEC at the connection points. $q_{si}$ can be obtained by coordinate transformation from $q_i$. To this end, we can claim that the dynamic model of the hybrid wind-wave energy system is established and the corresponding load is explained. The remaining problem is to obtain the control-oriented model based on Eq. \eqref{eq1}.

\subsection{Control-oriented model}
First, we choose the system state variables as $X_1=q_1,X_2=\dot{q}_1,X_3=q_2,X_4=\dot{q}_2,X_5=q_3,X_6=\dot{q}_3,
X_7=q_4,X_8=\dot{q}_4,X_9=q_5,X_{10}=\dot{q}_5,X_{11}=q_6,X_{12}=\dot{q}_6$, the state-space model of the hybrid wind-wave energy system can be written as:
\begin{equation}
  \label{eq11}
  \begin{cases}\dot{X}(t)&=\mathcal{A}X(t)+\mathcal{B}U(t)+\mathcal{H}\\Y(t)&=C^TX(t)\end{cases}
\end{equation}
where $U(t)=[ 0,F_{1}^{exe}(t),0,F_{2}^{exe}(t),0,F_{3}^{exe}(t),0,F_{4}^{exe}(t),0,$ $
F_{5}^{exe}(t),0,F_{6}^{exe}(t) ] ^T\in \mathbb{R}^{m\times 1}$ and $Y(t)=[ 1,0,1,0,1,0,1,0,1,$ $0,1,0 ] ^T\in \mathbb{R}^{m\times 1}$, where $m=12$. The system matrices $\mathcal{A}=[\mathcal{A}_1,\mathcal{A}_2;\mathcal{A}_3,\mathcal{A}_4]$ $\in\mathbb{R}^{m\times m}$, $\mathcal{B}=diag(0,s_3,0,p_5,0,h_3,0,l_7,$ $0,j_7,0,w_7)\in\mathbb{R}^{m\times m}$ and the constant vector $\mathcal{H}=[0_{5\times 1}, h_4, $ $0_{6\times 1}]\in\mathbb{R}^{m\times 1}$ ,where details of $\mathcal{A}_2=0\in \mathbb{R}^{\frac{m}{2}\times \frac{m}{2}}$ and $\mathcal{A}_i (i=1,3,4)$ are given as follows:
\begin{equation}\label{eq12}
\begin{aligned}
& \mathcal{A}_1 = \begin{bmatrix}
0 & \!1 & \!0 & \!0 & \!0 & \!0 \\
-s_1 & \!-s_2 & \!0 & \!0 & \!0 & \!0 \\
0 & \!0 & \!0 & \!1 & \!0 & \!0 \\
-p_3 & \!-p_4 & \!-p_1 & \!-p_2 & \!0 & \!0 \\
0 & \!0 & \!0 & \!0 & \!0 & \!1 \\
0 & \!0 & \!0 & \!0 & \!\!-h_1 & \!\!-h_2 
\end{bmatrix},
\mathcal{A}_3 = \begin{bmatrix}
0 & \!0 & \!0 & \!0 & \:0 & \:0 \\
-l_1 & \!-l_2 & \!-l_3 & \!-l_4 & \:0 & \:0 \\
0 & \!0 & \!0 & \!0 & \:0 & \:0 \\
-j_1 & \!-j_2 & \!-j_3 & \!-j_4 & \:0 & \:0 \\
0 & \!0 & \!0 & \!0 & \:0 & \:0 \\
-w_1 & \!-w_2 & \!-w_3 & \!-w_4 & \:0 & \:0 
\end{bmatrix}, \\
& \mathcal{A}_4 = \begin{bmatrix}
0 & \!1 & \!0 & \!0 & \!0 & \!0 \\
-l_5 & \!-l_6 & \!0 & \!0 & \!0 & \!0 \\
0 & \!1 & \!0 & \!0 & \!0 & \!0 \\
0 & \!0 & \!-j_5 & \!-j_6 & \!0 & \!0 \\
0 & \!1 & \!0 & \!0 & \!0 & \!0 \\
0 & \!0 & \!0 & \!0 & \!\!\!-w_5 & \!\!\!-w_6 
\end{bmatrix}
\end{aligned}
\end{equation}

In Eq. \eqref{eq12} $s_1=R_1/M_1$, $s_2=B_1/M_1$ and $s_3=1/M_1$. Similarly, $p_i$, $h_i$, $l_i$, $j_i$ and $w_i(i=1,2,3)$ have the same form as $s_i$. $p_4=B_{21}/M_{2}$, $p_5=1/M_{2}$ and $l_5=R_{42}/M_{4},\; l_6=B_{42}/M_{4}, \;l_7=1/M_{4}$, $j_i(i=5,6,7)$ and $w_i(i=5,6,7)$ can be expressed in the same form like $l_i$. Due to the presence of gravity in the heave mode as well as the pre-tension of the mooring system, a constant $h_4$ is set in the model. Note that all of the above parameters are unknown and the lumped external force $F_i^{exe}$ can be obtained from AQWA directly.

In order to facilitate the estimation of unknown parameters, Eq. \eqref{eq11} can be reformulated as the following linearly parameterized form:
\begin{equation}
  \label{eq13}
\dot{{X}}={\theta}{\Phi}(X)
\end{equation}
where ${\Phi}(X)=[X(t), U(t), 1]^T\in \mathbb{R}^{(m+m+1)\times 1}$ and $\theta=[\mathcal{A}, \mathcal{B},\mathcal{H}]\in \mathbb{R}^{m\times (m+m+1)}$.

\section{Adaptive parameter estimation algorithm}
In order to estimate the unknown parameters and achieve guaranteed convergence, we first define a set of low-pass filtered variables $X_{f}$ and $\Phi_{f}$ of $X$, $\Phi$ as:
\begin{equation}
  \label{eq14}
\left.\left\{\begin{aligned}&k\dot{X}_{f}+X_{f}=X,\qquad &x_{f}(0)=
0\\&k\dot{\Phi}_{f}+\Phi_{f}=\Phi,&\Phi_{f}(0)=0\end{aligned}\right.\right.
\end{equation}
where $k>0$ is the filter coefficient. Based on the first equation of Eq. \eqref{eq14}, $\dot{X}_f$ can be written along with Eq. \eqref{eq13} as follows:
\begin{equation}
  \label{eq15}
\dot{X}_{f}=\frac{X-X_{f}}{k}=\theta\Phi_{f}
\end{equation}

Then, two auxiliary matrices ${P}$ and ${Q}$ are defined as follows: 
\begin{equation}
  \label{eq16}
\left.\left\{\begin{aligned}\dot{P}&=-\ell P+\Phi_{f}\Phi_{f}^T,&P(0)=0\\\dot{Q}&=-\ell Q+[(X-X_{f})/k]\Phi_{f}^T,&Q(0)=0\end{aligned}\right.\right.
\end{equation}
where $\ell>0$ is the forgetting factor used to ensure the boundedness of $P$ and $Q$.  

By integrating on both sides of Eq.~(\ref{eq16}), the solution of $P$ and $Q$ can be computed as:
\begin{equation}
  \label{eq17}
\left.\left\{\begin{array}{l}P(t)\!=\!
\int_0^te^{-\ell(t-\tau)}\Phi_{f}(\tau)\Phi_{f}^T(\tau)d\tau\\\\Q(t)\!=\!
\int_0^te^{-\ell(t-\tau)}[(X(\tau)\!-\!X_{f}(\tau))/k]\Phi_{f}^T(\tau)d\tau\end{array}\right.\right.
\end{equation}

Then, another auxiliary matrix $W(t)\in \mathbb{R}^{m\times (m+m+1)}$ is designed based on $P(t)$ and $Q(t)$ as follows:
\begin{equation}
  \label{eq18}
W(t)=\hat{\theta}P(t)-Q(t)=-\tilde{\theta}P(t)
\end{equation}
where $\hat{\theta}$ is the estimation of the unknown parameter $\theta$ and $\tilde{\theta}=\theta-\hat{\theta}$ denotes the parameter estimation error. 

Now, the adaptive law is designed as:
\begin{equation}
  \label{eq19}
\dot{\hat{\theta}}=-\Gamma W
\end{equation}    
where $\Gamma$ is a design learning gain. 

To prove the convergence of the estimated parameter $\hat{\theta}$, we design the Lyapunov function candidate as $V=1/2\tilde{\theta}^T\Gamma^{-1}\tilde{\theta}$. Then, the time derivative of $V$ can be calculated along with Eq. \eqref{eq19} as:
\begin{equation}
  \label{eq20}
\dot{V}=\tilde{\theta}^T\Gamma^{-1}\dot{\tilde{\theta}}=
\tilde{\theta}^T{W}=-\tilde{\theta}^T{P}(t)\tilde{\theta}\leq-{\mu} V
\end{equation}
where $\mu=2\sigma/\lambda_{\max}(\Gamma^{-1})$ is a positive constant with $\lambda_{\max}\left(\cdot\right)$ being the maximum eigenvalue. According to the Lyapunov theory, we can conclude that the estimation error $\tilde{\theta}$ will converge to zero exponentially. 

According to the above mathematical developments, we can find that the proposed adaptive parameter estimation algorithm is driven by the parameter estimation error $\tilde{\theta}$. This is significantly different from the conventional parameter estimation methods, such as gradient descent method \cite{mishra2021analysis} and RLS \cite{li2024highly}. Recalling the conventional parameter estimation method, the predictor or observer is required to provide the state estimation error to drive the adaptive law. However, the parameter estimation error convergence cannot be guaranteed all the time as the estimation error is generally coupled with the predictor or observer error. If the system encounters with the high-frequency disturbance, such methods may emerge the well-known bursting phenomenon, which may trigger the system instability. From Eqs. \eqref{eq14}-\eqref{eq19}, one can find that only low-pass filter operation and  first order non-homogeneous differential equation are used to design the adaptive law Eq. \eqref{eq19}. In this respect, we can claim that the proposed adaptive parameter estimation algorithm is more straightforward as it is used in practice. Moreover, from the Lyapunov analysis, we can find that the adaptive law holds the exponential convergence characteristic, such that the parameter convergence can be guaranteed all the time.

Note that the true value of parameter $\theta$ is unknown, hence we cannot verify the effectiveness of the adaptive law by comparing the difference between the true value of parameter $\theta$ and estimated parameter $\hat{\theta}$. Inspired by literature \cite{Peng2023Optimization}, we conduct a BEM-based model of the hybrid wind-wave energy system in AQWA as the counterpart of the designed control-oriented model Eq. \eqref{eq11} with adaptive law Eq. \eqref{eq19}. By comparing the system responses of such two models, we can deduce if the estimated parameter converges to its true value. This process will be further clarified in simulation part.

\section{Numerical simulation and result analysis}
In this section, numerical simulation and performance analysis are conducted. The AQWA-Matlab simulation framework is established and the simulation setting in AQWA and environment conditions are introduced. Moreover, the gradient descent method is also studied in this section and chosen as the counterpart of the proposed adaptive parameter estimation algorithm to illustrate the superiority of the proposed method. 

\subsection{AQWA-Matlab simulation framework}
The AQWA-Matlab simulation framework is given in Fig. 2, where the data stream transferred between AQWA and Matlab can be observed. A BEM-based model as similar as \cite{Peng2023Optimization} is constructed in the AQWA and two wind and wave conditions in the North Sea \cite{HDCB202304001}, scenario 1: steady wind with regular wave and scenario 2: turbulent wind with irregular wave, are considered to obtain the system response and system load data. One can find from Fig. 2 that the yielded data from AQWA will serve as input for the Matlab port. Considering the unknown parameters involved into the control-oriented model of the hybrid wind-wave energy system, the proposed adaptive parameter estimation algorithm is implemented in Matlab. It is noteworthy that the proposed method uses the system response and system load data obtained from AQWA to estimate the unknown parameters in the model, and then substitute the estimated parameters back into the proposed control-oriented model to make a comparison to AQWA. 

\begin{figure*}[!htb]
	\centering
	\includegraphics[width=16cm]{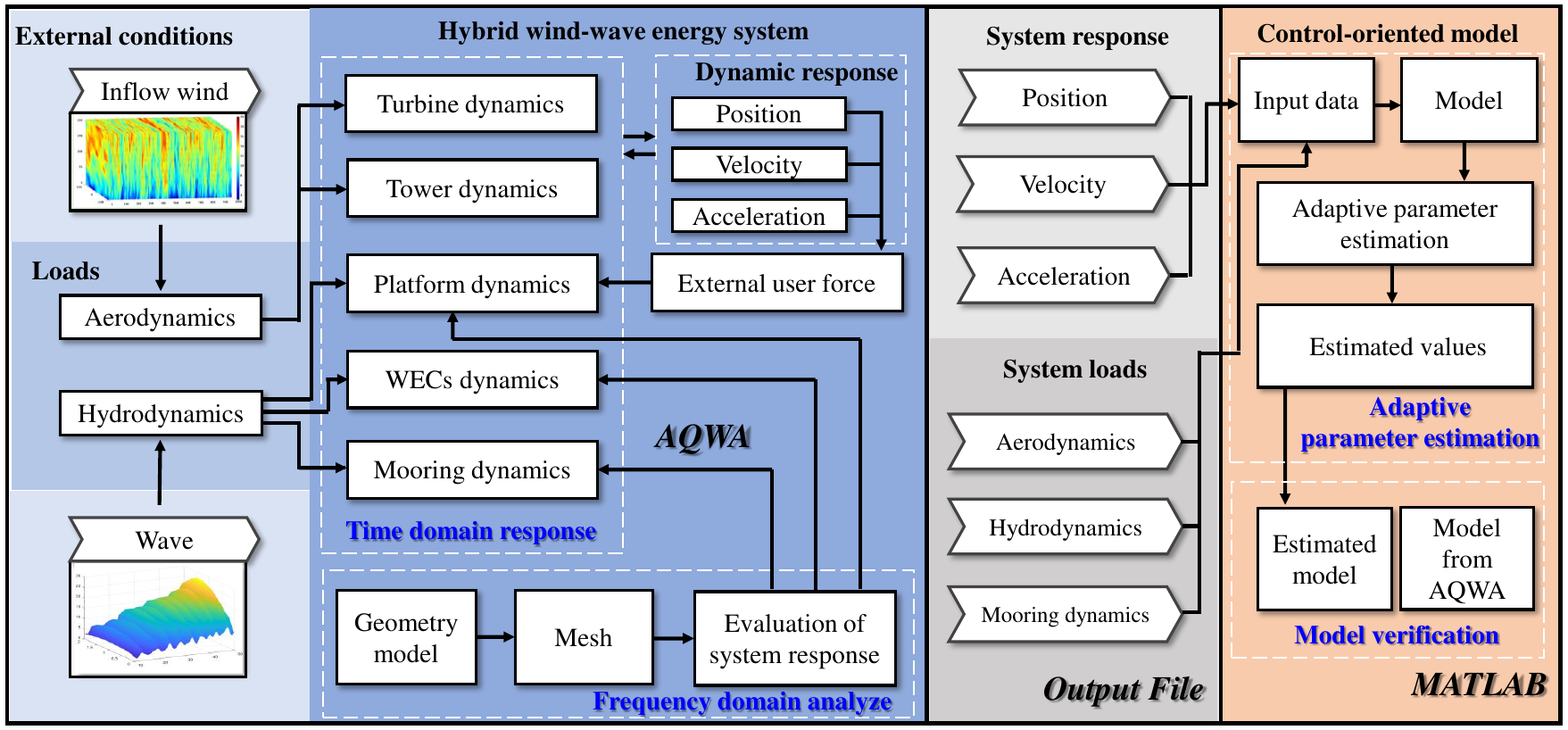}
	\caption{\fontfamily{ptm}\selectfont AQWA-Matlab framework}
	\label{fig2}
\end{figure*}

\subsection{Simulation setup}
The key parameters of the hybrid wind-wave energy system have been given in Table~\ref{tab1} and Table~\ref{tab2}, and the BEM-based model of the hybrid energy system is modelled in the environment of AQWA Workbench. The total number of mesh nodes generated is 26,211, and the maximum element size is $0.8~\mathrm{m}$. The system is anchored using a mooring system in a water area measuring $2000\times2000$ m with a depth of $200$ m. Additionally, three WECs are connected to the platform's offset columns using hinge joints, the PTO damping selected as $B_{PTO}=2e8~\mathrm{Nms}$ and the PTO stiffness is taken as zero, i.e., $K_{PTO}=0$, which can achieve the highest power production of the system \cite{Ghafari2021Numerical}. For the simulation in AQWA, the time duration is 200s and the time step is setting as $0.01~\mathrm{s}$.

Referring to the relevant wind and wave conditions of the Statfjord in the North Sea \cite{HDCB202304001}, the Joint
North Sea Wave Project (JONSWAP)  wave spectrum (peak factor, $\gamma=3.3$) is selected to generate irregular waves. Note that the regular wave can be generated by AQWA with the given significant wave height and peak period. Six typical operating sea conditions with the combined distribution of turbulent wind and irregular waves are selected, covering below rated power generation (Case 1, Case 2), rated power (Case 3), above rated power (Case 4) and extreme sea conditions above rated power (Case 5, Case 6). The adopted wind and wave parameters of six different operating sea conditions are depicted in Table~\ref{tab3}.
The simulation parameters used in Matlab are set as $x_i(0)=[0,0,0,0,-10.71,0]^T$, $\theta(0)=0$, $k=0.1$, $\ell=0.001$, $L=10$, the learning gain is $\Gamma=diag([0.23,0.09,0.12,221,500,$ $8,310,10,10,200,2,0.2,200])$, the learning gain is $G=diag([1000,200,1,10,$ $10,10,10,10,10,0.1,1000,1,10])$.

\begin{table}[htb]
  \centering
  \caption{Wind and wave parameters of six different operating sea conditions.}\label{tab3}
\tabcolsep=0.75cm \renewcommand\arraystretch{1.1}
  \begin{tabular}{@{}lllllll@{} }
\toprule
Case no. & $v\:(\mathrm{m/s})$  & $I\:(\mathrm{\%})$ & $H_s\:(\mathrm{m})$ & $T_p\:(\mathrm{s})$ & Power conditions & WEC status\\
\midrule
1 & 5   & 22.4 & 2.10 & 9.74 & Below rated power & Operational\\
2 & 10 & 15.7 & 2.88 & 9.98 & Below rated power & Operational\\
3 & 11.4 & 13.8 & 3.62 & 10.29 & Rated power & Operational\\
4 & 18 & 12.7 & 4.44 & 10.66 & Above rated power & Parked\\
5 & 22 & 12.1 & 5.32 & 11.06 & Above rated power & Parked\\
6 & 25 & 11.7 & 6.02 & 11.38 & Above rated power & Parked\\
\bottomrule
\end{tabular}
\end{table}

\subsection{Comparative simulation}
To demonstrate the superiority of the proposed parameter estimation method over conventional parameter estimation method,  the gradient descent method is adopted as the counterpart of the proposed parameter estimation method. First, a general predictor for system (\ref{eq13}) is designed as follows:
\begin{equation}
  \label{eq21}
\dot{\hat{x}}=\hat{\theta}\Phi (x)+L\left( x-\hat{x} \right) 
\end{equation}
where $L$ is the observer gain, $\hat{\theta}$ represents the estimated parameters to be online updated by the following adaptive law. Define the observer error as $e=x-\hat{x}$, then we can verify from Eq.~(\ref{eq13}) and Eq.~(\ref{eq21}) that:
\begin{equation}
  \label{eq22}
\dot{e}=\tilde{\theta}\Phi (x)-Le 
\end{equation}

The gradient descent adaptive law is formulated to minimize the estimation error $e$ (i.e., $\partial e^2/\partial \theta =0$), leading to an adaptive law by using the observer error $e$ as:
\begin{equation}
  \label{eq23}
\dot{\hat{\theta}}=G\frac{e}{n^2}\varPhi ^T 
\end{equation}
where $n^2$ is a normalizing factor given by $n=\sqrt{1+\varPhi ^{\mathrm{T}}\varPhi}$, and $G>0$ is the learning gain.

\subsection{Results analysis}
\subsubsection{Scenario 1: steady wind with regular wave}
Scenario 1 provides the comparative simulation results between the control-oriented model and BEM-based model under the steady wind with regular wave ocean environment. The parameter of steady wind is chosen as $v_{w}=5\mathrm{m/s}$ and the significant wave height and peak period for regular waves are chosen as $H_{s}=2\mathrm{m}$ and $T_{p}=11\mathrm{s}$, respectively. The wind and wave directions are chosen as $60^\circ$. The comparative system responses of the hybrid wind-wave energy system are provided in Fig. 3. As shown in Fig. 3, compared with the gradient descent method, the system responses of the platform on surge, pitch and heave modes, the WECs pitch of the proposed method show good agreement with AQWA. This result indicates that the proposed model can effectively predict the system dynamics in regular wave and steady state wind environments. Notably, this data was used as the data source for adaptive parameter estimation algorithm and the estimated parameters were applied to the system to observe the system response under other environmental conditions later. The obtained good agreement about the system responses can be further explained by Fig. 4, in which the estimation results of the unknown parameters of the hybrid energy system with gradient descent and the proposed parameter estimation method are depicted. It can be seen from Fig. 4 that the proposed parameter estimation method can guarantee the convergence of the estimated parameters, i.e., less fluctuations can be found. Conversely, the aggressive fluctuations encountered into unknown parameter $s_1, p_5$ using gradient descent adaptive law, whilst it cannot drive the unknown parameters to their true values. This can be attributed to the fact that the system responses of the control-oriented model in Fig. 3 cannot achieve good agreement in comparison to the system responses with the proposed parameter estimation method. Considering the length of this paper, we present all the estimated parameters of the platform and WECs in Table~\ref{tab4} instead of figure. Note that we take the average values of the estimated parameters in the last ten seconds to eliminate the effect of parameter fluctuation. In Table~\ref{tab4}, the upper rows represent the parameter estimation with proposed method and the lower rows represent the parameter estimation with the gradient descent method.

\begin{figure}[!htb]
	\centering
	\includegraphics[width=17.5cm]{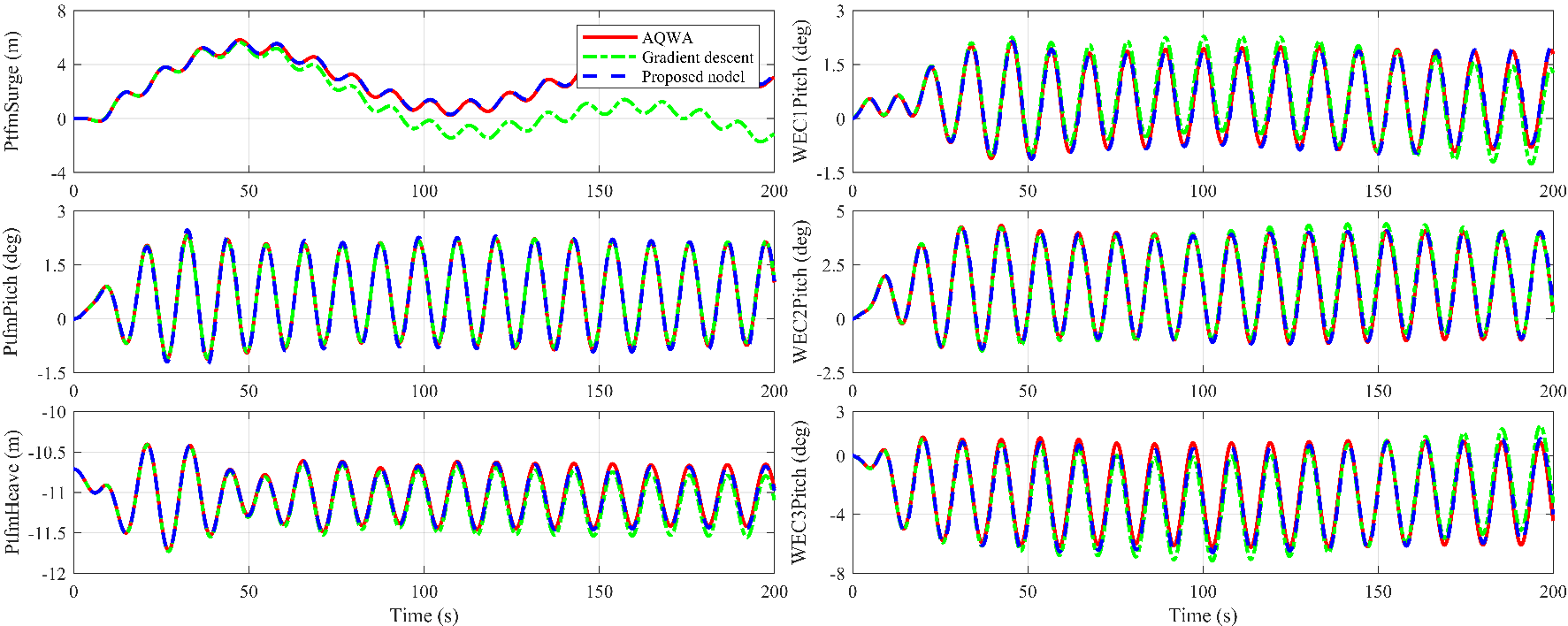}
	\caption{\fontfamily{ptm}\selectfont Dynamic response comparison between AQWA and proposed model with $60^\circ$ wind-wave direction ($5 \: m/s$ steady wind, regular wave with $1.5\: m\: H_s$ and $10\: s\: T_p$).}
    	\label{fig3}
\end{figure}

\begin{figure*}[!htb]
	\centering
	\begin{minipage}[t]{0.33\linewidth}
		\centering
	\includegraphics[width=\hsize]{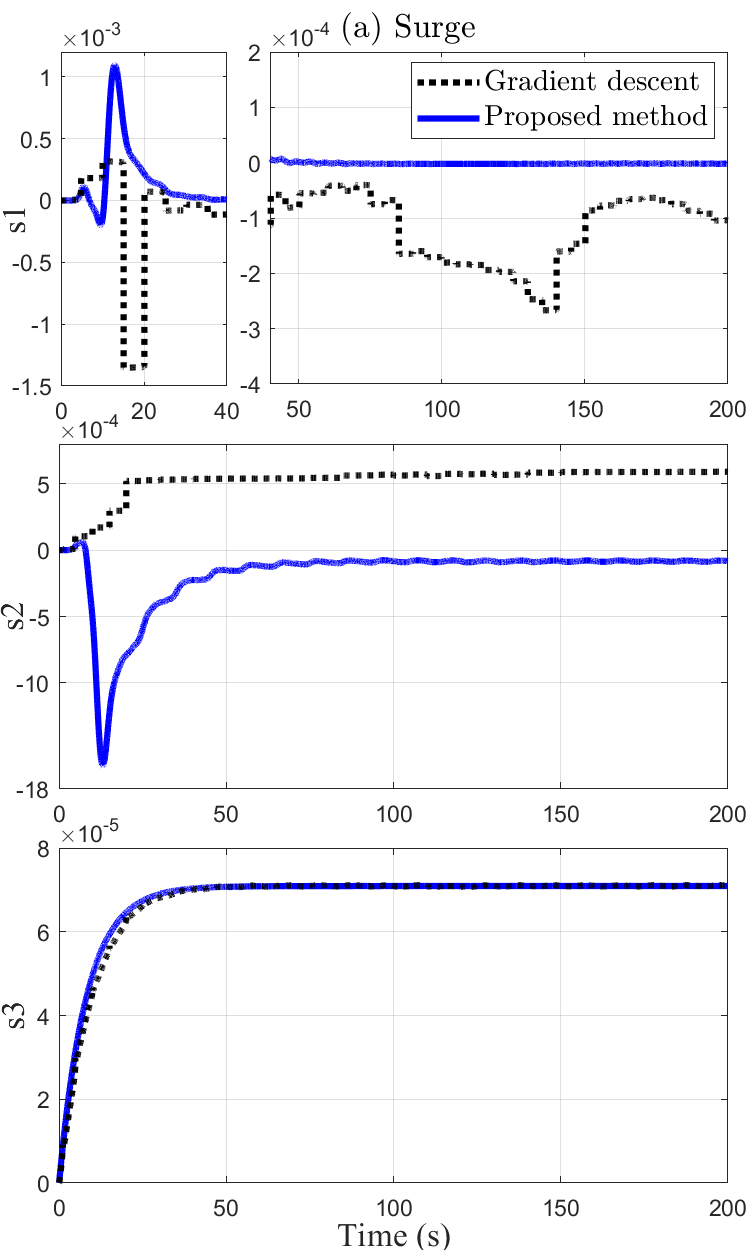}\\
	\end{minipage}%
	\hfill
	\begin{minipage}[t]{0.33\linewidth}
		\centering
	\includegraphics[width=\hsize]{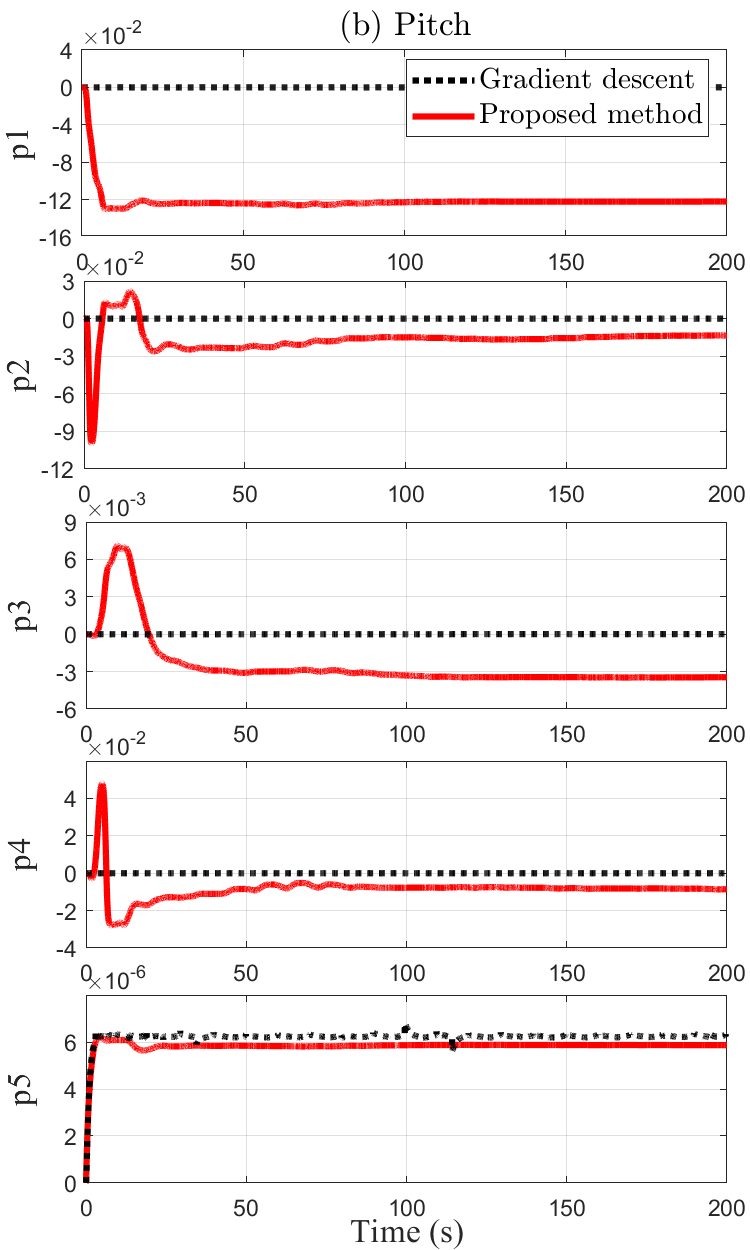}\\
	\end{minipage}%
	\begin{minipage}[t]{0.33\linewidth}
		\centering
	\includegraphics[width=\hsize]{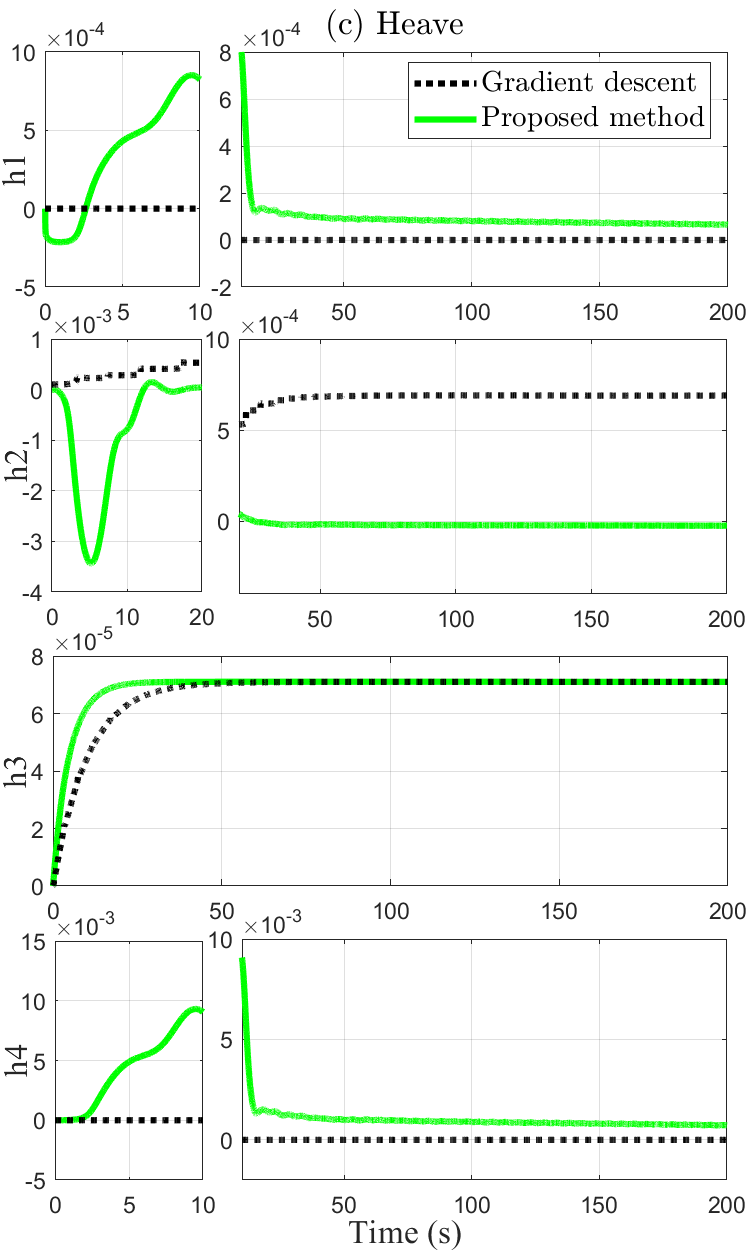}\\
	\end{minipage}%
  \caption{\fontfamily{ptm}\selectfont Parameter estimation with adaptation law ~(\ref{eq19}) and gradient descent method~(\ref{eq23}). (a) Surge. (b) Pitch. (c) Heave.}
  \label{fig4}
\end{figure*}

\begin{table}[htb]
  \centering
  \caption{Estimated values of system parameters with adaptation law ~(\ref{eq19}) and gradient descent method~(\ref{eq23}).}\label{tab4}
  \tiny
\tabcolsep=0.33cm \renewcommand\arraystretch{1.1}
  \begin{tabular}{@{} llllllll@{} }
    \toprule
        \fontfamily{ptm}\selectfont Mode\&Method & \fontfamily{ptm}\selectfont Parameters\&Values & ~ & ~ & ~ & ~ & ~ & ~ \\ 
    \midrule
        \fontfamily{ptm}\selectfont surge & \fontfamily{ptm}\selectfont s1 & \fontfamily{ptm}\selectfont s2 & \fontfamily{ptm}\selectfont s3 & ~ & ~ & ~ & ~ \\ 
        
        \fontfamily{ptm}\selectfont Eq.~(\ref{eq19}) & \fontfamily{ptm}\selectfont $-5.827\times10^{-7}$ & \fontfamily{ptm}\selectfont $-6.314\times10^{-5}$ & \fontfamily{ptm}\selectfont $7.092\times10^{-5}$ & ~ & ~ & ~ & ~ \\ 
        
        \fontfamily{ptm}\selectfont Eq.~(\ref{eq23}) & \fontfamily{ptm}\selectfont $-1.135\times10^{-4}$ & \fontfamily{ptm}\selectfont $1.292\times10^{-4}$ & \fontfamily{ptm}\selectfont $7.097\times10^{-5}$ & ~ & ~ & ~ & ~ \\ 
        \midrule
        \fontfamily{ptm}\selectfont pitch & \fontfamily{ptm}\selectfont p1 & \fontfamily{ptm}\selectfont p2 & \fontfamily{ptm}\selectfont p3 & \fontfamily{ptm}\selectfont p4 & \fontfamily{ptm}\selectfont p5 & ~ & ~ \\ 
        
        \fontfamily{ptm}\selectfont Eq.~(\ref{eq19}) & \fontfamily{ptm}\selectfont $-0.126$ & \fontfamily{ptm}\selectfont -0.186 & \fontfamily{ptm}\selectfont -0.003 & \fontfamily{ptm}\selectfont -0.009 & \fontfamily{ptm}\selectfont $-5.868\times10^{-6}$ & ~ & ~ \\ 
        
        \fontfamily{ptm}\selectfont Eq.~(\ref{eq23}) & \fontfamily{ptm}\selectfont $7.488\times10^{-7}$ & \fontfamily{ptm}\selectfont $-4.921\times10^{-8}$ & \fontfamily{ptm}\selectfont $1.119\times10^{-6}$ & \fontfamily{ptm}\selectfont $-5.897\times10^{-8}$ & \fontfamily{ptm}\selectfont $6.149\times10^{-6}$ & ~ & ~ \\ 
        \midrule
        \fontfamily{ptm}\selectfont heave & \fontfamily{ptm}\selectfont h1 & \fontfamily{ptm}\selectfont h2 & \fontfamily{ptm}\selectfont h3 & \fontfamily{ptm}\selectfont h4 & ~ & ~ & ~ \\ 
        
        \fontfamily{ptm}\selectfont Eq.~(\ref{eq19}) & \fontfamily{ptm}\selectfont  $-6.363\times10^{-5}$& \fontfamily{ptm}\selectfont $-2.351\times10^{-5}$ & \fontfamily{ptm}\selectfont $7.102\times10^{-5}$ & \fontfamily{ptm}\selectfont $7.029\times10^{-4}$ & ~ & ~ & ~ \\ 
        
        \fontfamily{ptm}\selectfont Eq.~(\ref{eq23}) & \fontfamily{ptm}\selectfont $1.205\times10^{-7}$ & \fontfamily{ptm}\selectfont $6.896\times10^{-4}$ & \fontfamily{ptm}\selectfont $7.101\times10^{-5}$ & \fontfamily{ptm}\selectfont $-1.109\times10^{-6}$ & ~ & ~ & ~ \\ 
        \midrule
        \fontfamily{ptm}\selectfont WEC1 & \fontfamily{ptm}\selectfont l1 & \fontfamily{ptm}\selectfont l2 & \fontfamily{ptm}\selectfont l3 & \fontfamily{ptm}\selectfont l4 & \fontfamily{ptm}\selectfont l5 & \fontfamily{ptm}\selectfont l6 & \fontfamily{ptm}\selectfont l7 \\ 
        
        \fontfamily{ptm}\selectfont Eq.~(\ref{eq19}) & \fontfamily{ptm}\selectfont $-1.530\times10^{-3}$ & \fontfamily{ptm}\selectfont $-2.400\times10^{-3}$ & \fontfamily{ptm}\selectfont $-6.878\times10^{-6}$ & \fontfamily{ptm}\selectfont $7.173\times10^{-4}$ & \fontfamily{ptm}\selectfont $1.115\times10^{-3}$ & \fontfamily{ptm}\selectfont $1.272\times10^{-3}$ & \fontfamily{ptm}\selectfont $0.055$ \\ 
        
          \fontfamily{ptm}\selectfont Eq.~(\ref{eq23}) & \fontfamily{ptm}\selectfont $-4.025\times10^{-4}$ & \fontfamily{ptm}\selectfont $8.245\times10^{-4}$ & \fontfamily{ptm}\selectfont $3.849\times10^{-5}$ & \fontfamily{ptm}\selectfont $7.883\times10^{-5}$ & \fontfamily{ptm}\selectfont $4.933\times10^{-5}$ & \fontfamily{ptm}\selectfont $7.756\times10^{-3}$ & \fontfamily{ptm}\selectfont $0.055$ \\ 
        \midrule
        \fontfamily{ptm}\selectfont WEC2 & \fontfamily{ptm}\selectfont j1 & \fontfamily{ptm}\selectfont j2 & \fontfamily{ptm}\selectfont j3 & \fontfamily{ptm}\selectfont j4 & \fontfamily{ptm}\selectfont j5 & \fontfamily{ptm}\selectfont j6 & \fontfamily{ptm}\selectfont j7 \\ 
        
        \fontfamily{ptm}\selectfont Eq.~(\ref{eq19}) & \fontfamily{ptm}\selectfont $2.787\times10^{-4}$ & \fontfamily{ptm}\selectfont $2.004\times10^{-4}$ & \fontfamily{ptm}\selectfont $-2.388\times10^{-5}$ & \fontfamily{ptm}\selectfont $-8.288\times10^{-4}$ & \fontfamily{ptm}\selectfont $-6.739\times10^{-4}$ & \fontfamily{ptm}\selectfont $-1.779\times10^{-3}$ & \fontfamily{ptm}\selectfont $0.055$ \\ 
        
        \fontfamily{ptm}\selectfont Eq.~(\ref{eq23}) & \fontfamily{ptm}\selectfont  $-6.465\times10^{-4}$& \fontfamily{ptm}\selectfont $7.940\times10^{-4}$ & \fontfamily{ptm}\selectfont $7.378\times10^{-5}$ & \fontfamily{ptm}\selectfont $-1.468\times10^{-3}$ & \fontfamily{ptm}\selectfont $1.103\times10^{-3}$ & \fontfamily{ptm}\selectfont $-5.096\times10^{-4}$ & \fontfamily{ptm}\selectfont $0.056$ \\ 
        \midrule
        \fontfamily{ptm}\selectfont WEC3 & \fontfamily{ptm}\selectfont w1 & \fontfamily{ptm}\selectfont w2 & \fontfamily{ptm}\selectfont w3 & \fontfamily{ptm}\selectfont w4 & \fontfamily{ptm}\selectfont w5 & \fontfamily{ptm}\selectfont w6 & \fontfamily{ptm}\selectfont w7 \\ 
        
        \fontfamily{ptm}\selectfont Eq.~(\ref{eq19}) & \fontfamily{ptm}\selectfont $-2.892\times10^{-4}$ & \fontfamily{ptm}\selectfont $-1.436\times10^{-4}$ & \fontfamily{ptm}\selectfont $-2.082\times10^{-5}$ & \fontfamily{ptm}\selectfont $-4.999\times10^{-4}$ & \fontfamily{ptm}\selectfont $-8.686\times10^{-4}$ & \fontfamily{ptm}\selectfont $1.490\times10^{-4}$ & \fontfamily{ptm}\selectfont $0.055$ \\ 
        
        \fontfamily{ptm}\selectfont Eq.~(\ref{eq23}) & \fontfamily{ptm}\selectfont $-4.571\times10^{-4}$ & \fontfamily{ptm}\selectfont $5.779\times10^{-5}$ & \fontfamily{ptm}\selectfont $-1.146\times10^{-4}$ & \fontfamily{ptm}\selectfont $-9.820\times10^{-4}$ & \fontfamily{ptm}\selectfont $-7.007\times10^{-4}$ & \fontfamily{ptm}\selectfont $-7.738\times10^{-4}$ & \fontfamily{ptm}\selectfont $0.055$ \\ 
    \bottomrule
\end{tabular}
\end{table}

\subsubsection{Scenario 2: turbulent wind with irregular wave}
To further verify the fidelity of the proposed control-oriented model, six typical sea conditions with turbulent wind and irregular wave, as shown in Table~\ref{tab3}, are employed. Meanwhile, three different wind and wave directions, $30^\circ$, $60^\circ$ and $90^\circ$, are also considered in Scenario 2 to further demonstrate the higher fidelity of the proposed control-oriented model. The system responses of the control-oriented model under Case 3 sea condition with three wind-wave directions are given in Fig. 5. One can find from Fig. 5 that the proposed control-oriented model with the proposed parameter estimation method can accurately predict the system response of the AQWA in comparison to that of control-oriented model with gradient descent method. This fact further demonstrates that the proposed parameter estimation method contrisidewaystablebutes to obtaining a higher-fidelity control-oriented model.  
\begin{figure*}[!htb]
 \centering
	\includegraphics[width=17.5cm]{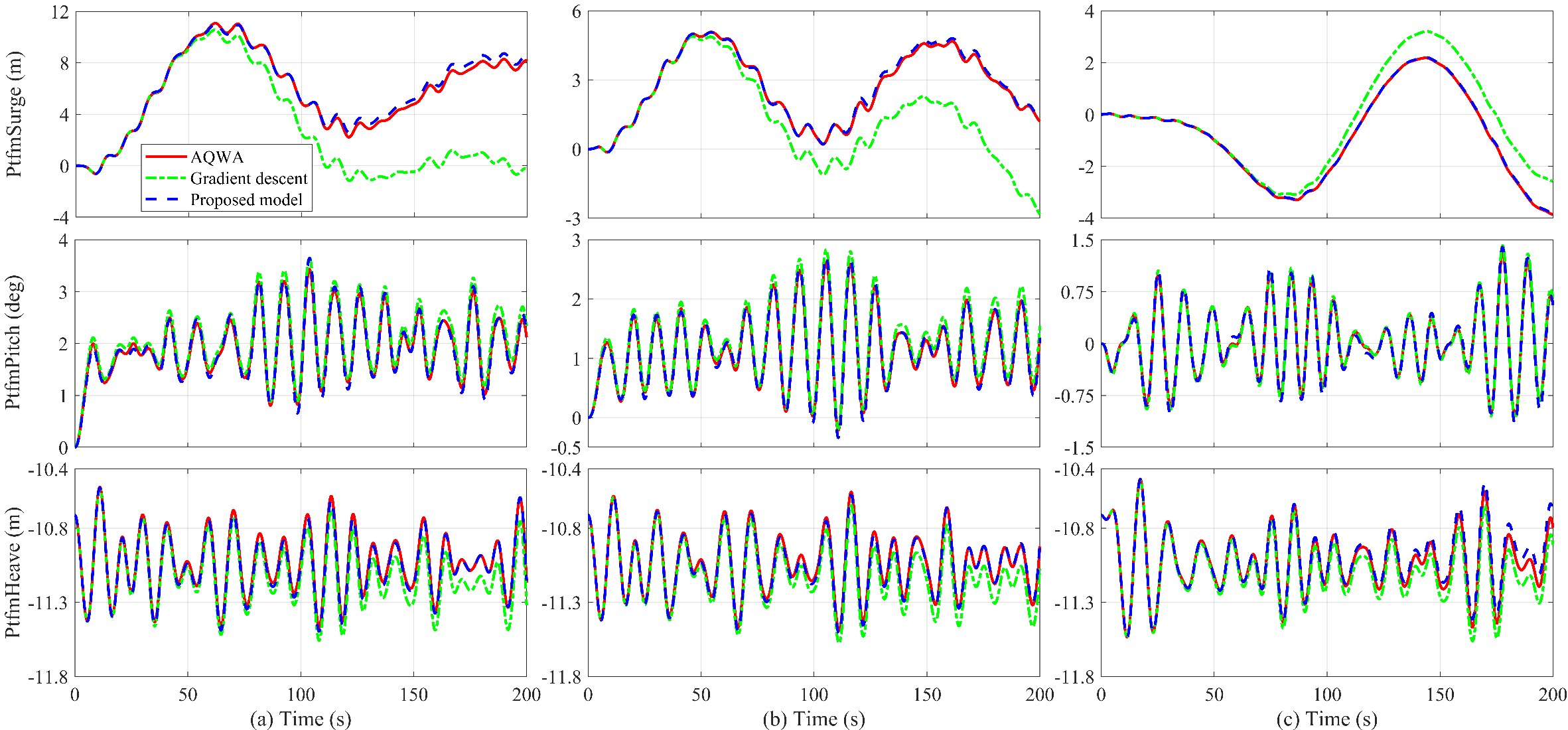}
	\caption{\fontfamily{ptm}\selectfont Dynamic response comparison between AQWA and proposed model with $30^\circ, 60^\circ$ and $90^\circ$ wind-wave direction ($11.4\: m/s$ turbulent wind, irregular wave with $3.62\: m\: H_s$ and $10.29\: s\: T_p$), (a) $30^\circ$, (b) $60^\circ$, (c) $90^\circ$.}
    	\label{fig5}
\end{figure*}

\begin{figure*}[!htb]
	\centering
	\begin{minipage}[t]{0.47\linewidth}
		\centering
		\includegraphics[width=\hsize]{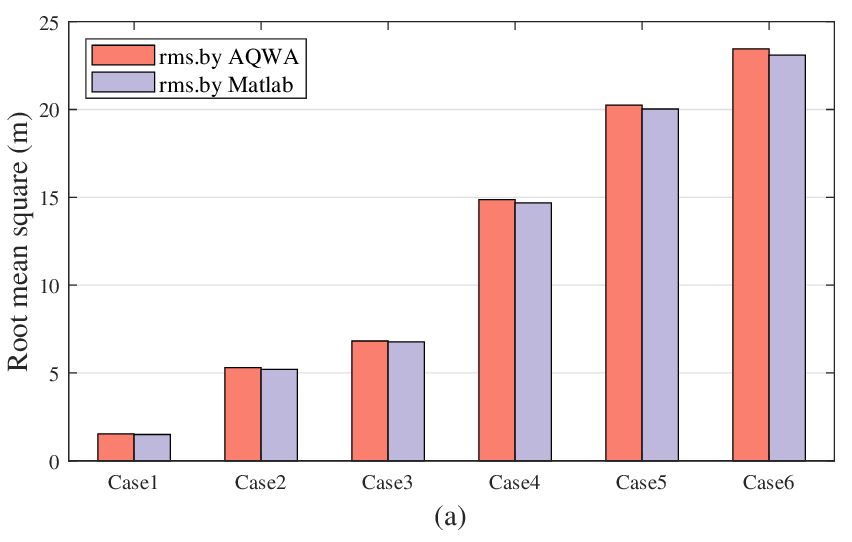}\\
	\end{minipage}%
	\hfill
	\begin{minipage}[t]{0.47\linewidth}
		\centering
		\includegraphics[width=\hsize]{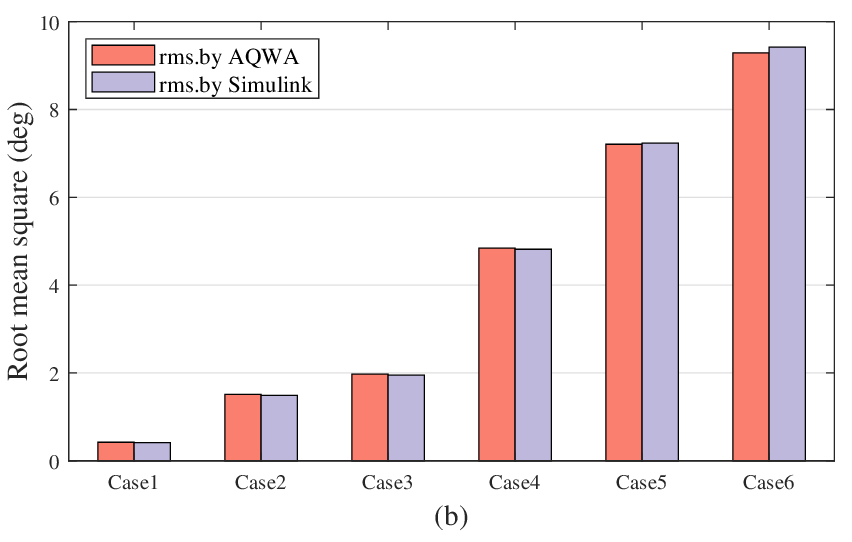}\\
	\end{minipage}%
	\begin{minipage}[t]{0.47\linewidth}
		\centering
		\includegraphics[width=\hsize]{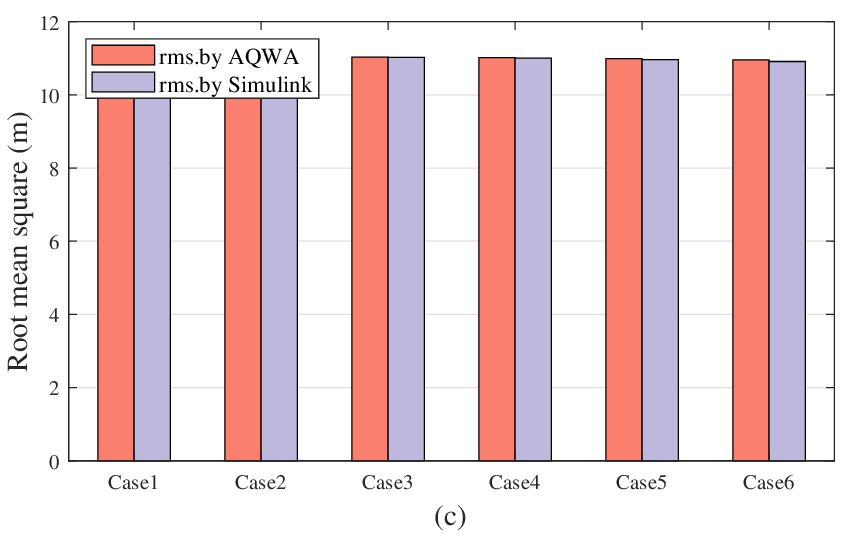}\\
	\end{minipage}%
	\caption{\fontfamily{ptm}\selectfont Root mean square comparison between AQWA and proposed model in six cases, (a) Surge, (b) Pitch, (c) Heave.}
	  \label{fig6}
\end{figure*}

\begin{figure*}[!htb]
	\centering
	\includegraphics[width=12cm]{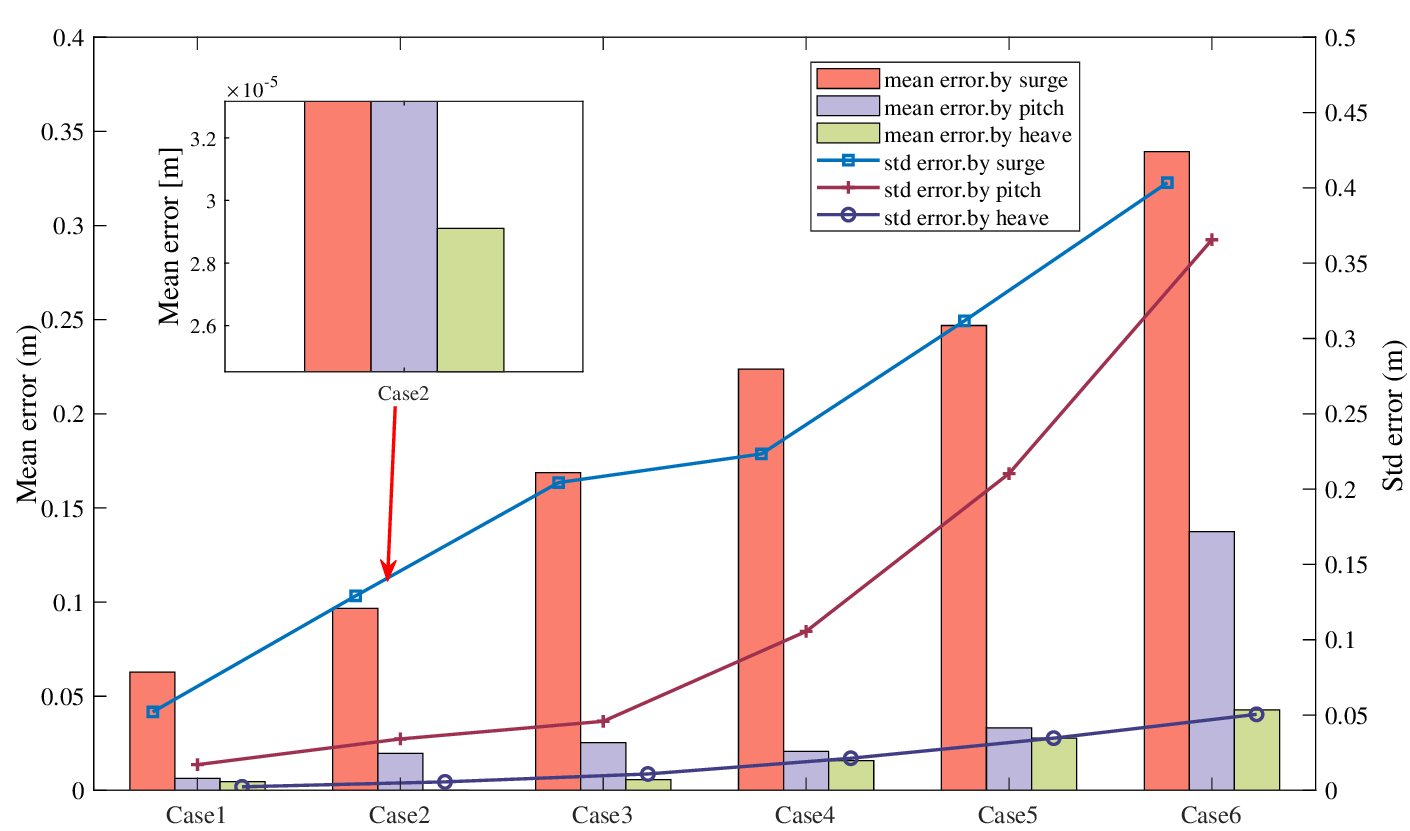}
	\caption{\fontfamily{ptm}\selectfont Standard deviation and mean of the error in six cases.}
    	\label{fig7}
\end{figure*}

\begin{figure*}[!htb]
	\centering
	\begin{minipage}[t]{0.47\linewidth}
		\centering
		\includegraphics[width=\hsize]{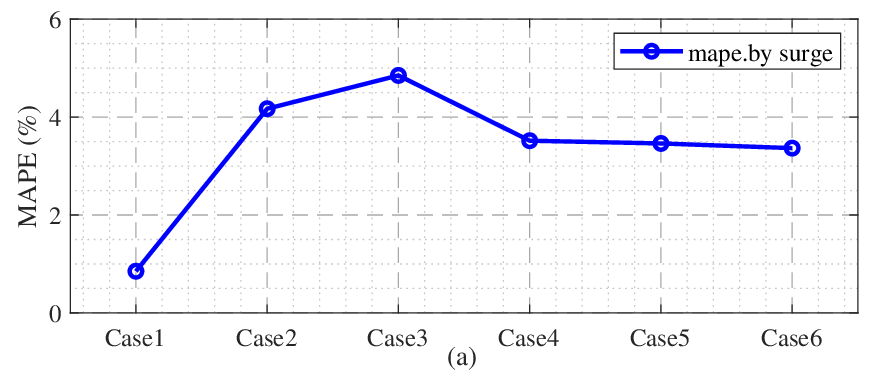}\\
	\end{minipage}%
	\hfill
	\begin{minipage}[t]{0.47\linewidth}
		\centering
		\includegraphics[width=\hsize]{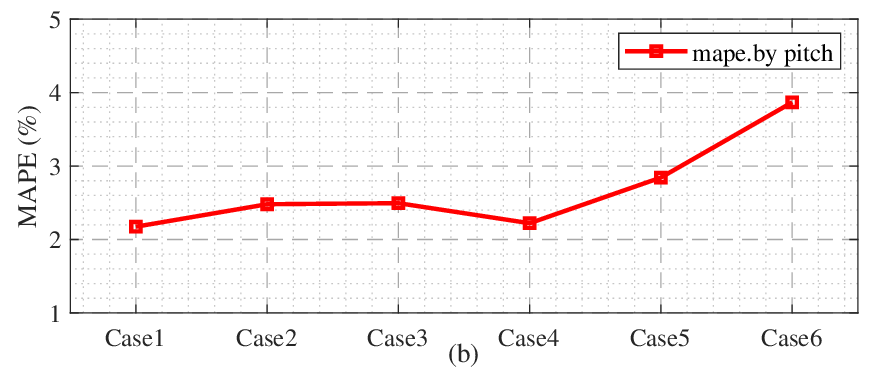}\\
	\end{minipage}%
	\begin{minipage}[t]{0.47\linewidth}
		\centering
		\includegraphics[width=\hsize]{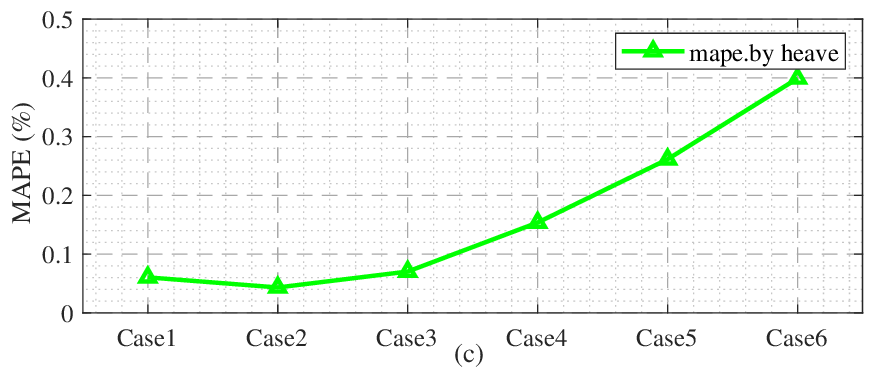}\\
	\end{minipage}%
  \caption{\fontfamily{ptm}\selectfont Mean Absolute Percentage Error of three nodes in six cases, (a) Surge, (b) Pitch, (c) Heave.}
	  \label{fig8}
\end{figure*}

Furthermore, to demonstrate the good robustness of the proposed control-oriented model, it is tested on six sea conditions along with wind and wave directions $30^\circ$. The statistical results are depicted in Figs. 6-9. In Fig. 6, the root mean square (RMS) performance index is introduced to evaluate the trends of the three motion modes. From Fig. 6(a) and Fig. 6(b), we can find that both the pitch and surge modes show an increasing trend as the sea condition gets worse in comparison to the heave mode. This is nature because heave mode is mainly affected by the vertical motion of ocean wave, but the wave frequency may not match the system's natural heave frequency. In contrast, both the surge and pitch modes are affected by the horizontal thrust exerted by wave and wind, which is more likely to cause significant motion. Moreover, among the three motion modes, aggressive pitch motion may reduce the device fatigue life and even damage the whole system. In this sense, developing a high fidelity control-oriented model contributes to designing the advanced hybrid wind-wave energy system control method and guaranteeing its safety when subjected to harsh sea conditions. The profiles of the standard deviation (STD) error and mean error with respect to the three modes of the control-oriented model are given in Fig. 7. As shown in Fig. 7, both the STD error and mean error increase as the sea condition gets worse. As discussed in \cite{zhang2020hydrodynamic} and parameter estimation \cite{anderson2005failures}, a signal with higher frequency is more likely to trigger the unmodeled dynamics, which can significantly degrade the accuracy of the established control-oriented model. This fact is inevitable. In this sense, the mean absolute percentage error (MAPE) is proposed and widely-accepted as performance index to evaluate the fidelity of the established control-oriented model. Similar to \cite{YANG2020Development}, we choose $5\%$ MAPE as the threshold of the fidelity of the control-oriented model and the corresponding results are given in Fig. 8. As shown in Fig. 8(a), the MAPE value reaches a peak of 4.86\% in Case 3, all MAPE value under different sea conditions are below $5\%$. Finally, the performance index r-square is introduced to evaluate the correlation between the proposed control-oriented model in Matlab and BEM-based model in AQWA. It is well-know that if the value of r-square get closer to 1, the higher correlation can be obtained between the proposed model and the benchmark model \cite{larsen2005introduction}. From Fig. 9, we can find that the r-square values for all the three motion modes are over 0.9 and almost close to 1. It is noteworthy that the r-square of the pitch motion in Case 6 holds the lowest value 0.95 than that of other motion modes. This fact is reasonable because the pitch mode is more sensitive to the harsh sea condition, indicating the emergence of the unmodeled dynamics and therefore degrade the model correlation between the control-oriented model and BEM-based model. Taking into account the aforementioned comparative simulation results, we can claim that the established control-oriented model Eq. \eqref{eq13} with the proposed parameter estimation method  Eq. \eqref{eq19} holds great agreement with the BEM-based model, indicating higher fidelity. 

\begin{figure*}[!htb]
\centering
	\includegraphics[width=13cm]{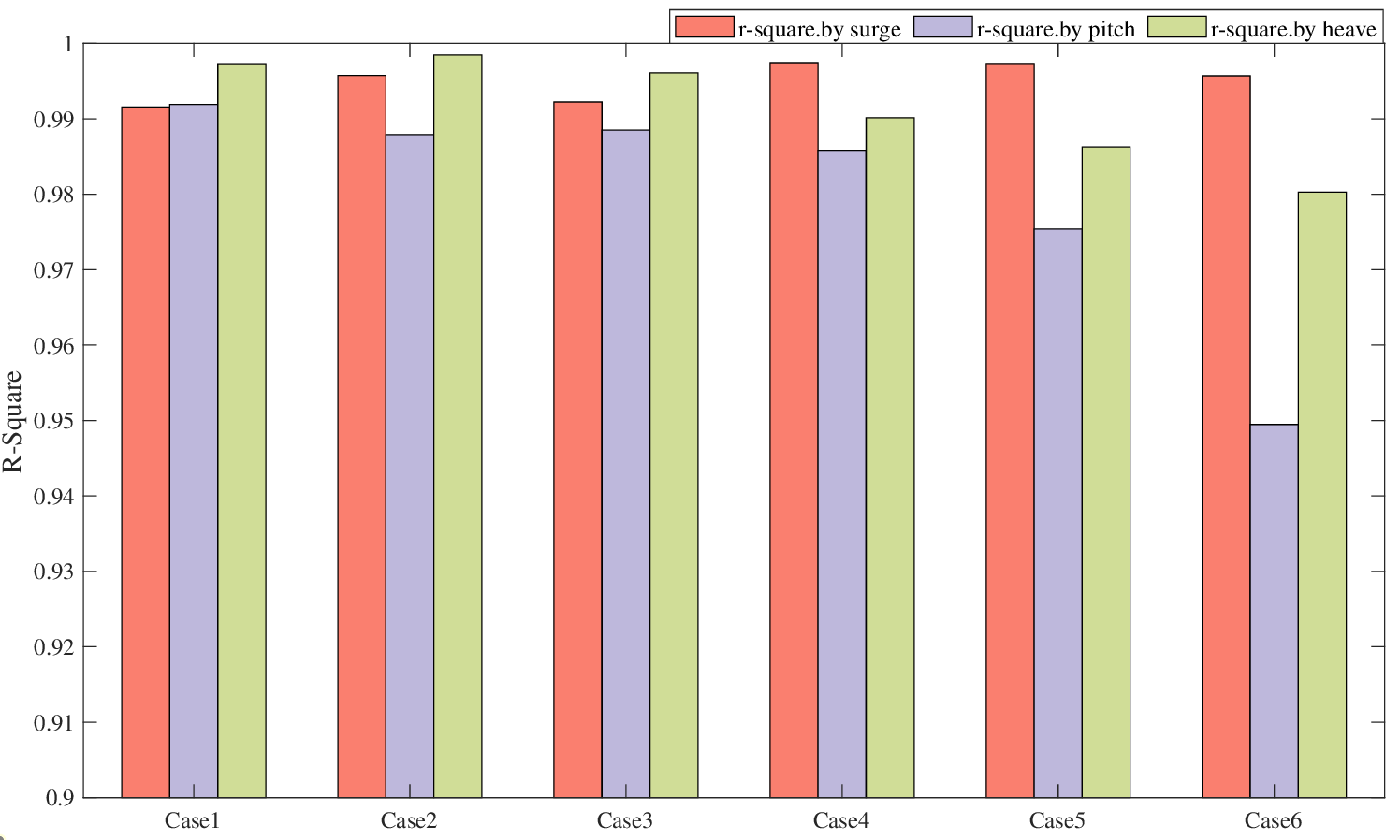}
	\caption{\fontfamily{ptm}\selectfont R-Square of the platform in three nodes (surge, pitch, heave).}
    	\label{fig9}
\end{figure*}

\section{Conclusion}
This paper proposed a control-oriented model for the hybrid wind-wave energy system. A new adaptive parameter estimation algorithm driven by parameter estimation error was proposed to address the system unknown parameter. Comparative numerical analyses concerning on motion response of the hybrid wind-wave energy system were conducted in terms of the designed AQWA-Matlab framework. The conclusions in this paper are presented below  
\begin{enumerate}[\textbullet]
\item The proposed control-oriented model of the hybrid wind-wave energy system can accurately predict the system motion response under various environmental conditions in comparison to the BEM-based model, demonstrating the good fidelity of the proposed model. The proposed model has the characteristics of low order, simple model structure and high precision. It effectively solves the issue that most existing models are not benefited to controller design.    
\item Different from the conventional parameter estimation methods (i. e., gradient descent method, RLS) driven by predictor and/or observer error, the proposed adaptive parameter estimation method does not require any predictor or observer technique. Moreover, the proposed method is driven by the parameter estimation error information, such that the potential bursting phenomenon can be remedied and the system unknown parameters can be driven to their true values. Moreover, the implementation of the proposed parameter estimation method only requires a set of low-pass filter operations and auxiliary matrices, such that the proposed method has simple structure and benefits to practical application.
\end{enumerate}

In future study, we will focus on eliminating the unmodeled dynamics when the control-oriented model suffers from harsh sea conditions. Also, control strategies based on the proposed model will be studied to mitigate the motion response of the platform, whilst improving the power generation efficiency.

\bibliographystyle{IEEEtran}
\bibliography{Reference}

\begin{thebibliography}{10}
\providecommand{\url}[1]{#1}
\csname url@samestyle\endcsname
\providecommand{\newblock}{\relax}
\providecommand{\bibinfo}[2]{#2}
\providecommand{\BIBentrySTDinterwordspacing}{\spaceskip=0pt\relax}
\providecommand{\BIBentryALTinterwordstretchfactor}{4}
\providecommand{\BIBentryALTinterwordspacing}{\spaceskip=\fontdimen2\font plus
\BIBentryALTinterwordstretchfactor\fontdimen3\font minus \fontdimen4\font\relax}
\providecommand{\BIBforeignlanguage}[2]{{%
\expandafter\ifx\csname l@#1\endcsname\relax
\typeout{** WARNING: IEEEtran.bst: No hyphenation pattern has been}%
\typeout{** loaded for the language `#1'. Using the pattern for}%
\typeout{** the default language instead.}%
\else
\language=\csname l@#1\endcsname
\fi
#2}}
\providecommand{\BIBdecl}{\relax}
\BIBdecl

\bibitem{shi2023real}
W.~Shi, J.~Fu, Z.~Ren, Z.~Jiang, T.~Wang, L.~Cui, and X.~Li, ``Real-time hybrid model tests of floating offshore wind turbines: Status, challenges, and future trends,'' \emph{Applied Ocean Research}, vol. 141, p. 103796, 2023.

\bibitem{Suzuki2007Load}
H.~Suzuki and A.~Sato, ``Load on turbine blade induced by motion of floating platform and design requirement for the platform,'' in \emph{International Conference on Offshore Mechanics and Arctic Engineering}, vol. 42711, 2007, pp. 519--525.

\bibitem{de2024assessment}
S.~P. De~Le{\'o}n, J.~H. Bettencourt, J.~V. Ringwood, and J.~Benveniste, ``Assessment of combined wind and wave energy in european coastal waters using satellite altimetry.'' \emph{Applied Ocean Research}, vol. 152, p. 104184, 2024.

\bibitem{perez2015review}
C.~P{\'e}rez-Collazo, D.~Greaves, and G.~Iglesias, ``A review of combined wave and offshore wind energy,'' \emph{Renewable and sustainable energy reviews}, vol.~42, pp. 141--153, 2015.

\bibitem{zhang2024computational}
W.~Zhang, J.~Calderon-Sanchez, D.~Duque, and A.~Souto-Iglesias, ``Computational fluid dynamics (cfd) applications in floating offshore wind turbine (fowt) dynamics: A review,'' \emph{Applied Ocean Research}, vol. 150, p. 104075, 2024.

\bibitem{nielsen2015partnership}
K.~Nielsen, J.~Krogh, H.~J. Brodersen, P.~Steenstrup, H.~Pilgaard, L.~Marquis, E.~Friis-Madsen, and J.~Kofoed, ``Partnership for wave power-roadmaps,'' Citeseer, Tech. Rep., 2015.

\bibitem{Peng2023Optimization}
P.~Jin, Z.~Zheng, Z.~Zhou, B.~Zhou, L.~Wang, Y.~Yang, and Y.~Liu, ``Optimization and evaluation of a semi-submersible wind turbine and oscillating body wave energy converters hybrid system,'' \emph{Energy}, vol. 282, p. 128889, 2023.

\bibitem{zhang2022coupled}
D.~Zhang, Z.~Chen, X.~Liu, J.~Sun, H.~Yu, W.~Zeng, Y.~Ying, Y.~Sun, L.~Cui, S.~Yang \emph{et~al.}, ``A coupled numerical framework for hybrid floating offshore wind turbine and oscillating water column wave energy converters,'' \emph{Energy Conversion and Management}, vol. 267, p. 115933, 2022.

\bibitem{Jonkman2005FAST}
J.~M. Jonkman and M.~L. Buhl~Jr, ``Fast user's guide-updated august 2005,'' National Renewable Energy Lab.(NREL), Golden, CO (United States), Tech. Rep., 2005.

\bibitem{si2021influence}
Y.~Si, Z.~Chen, W.~Zeng, J.~Sun, D.~Zhang, X.~Ma, and P.~Qian, ``The influence of power-take-off control on the dynamic response and power output of combined semi-submersible floating wind turbine and point-absorber wave energy converters,'' \emph{Ocean Engineering}, vol. 227, p. 108835, 2021.

\bibitem{Zhu2023Optimal}
H.~Zhu, ``Optimal semi-active control for a hybrid wind-wave energy system on motion reduction,'' \emph{IEEE Transactions on Sustainable Energy}, vol.~14, no.~1, pp. 75--82, 2022.

\bibitem{YANG2020Development}
Y.~Yang, M.~Bashir, C.~Michailides, C.~Li, and J.~Wang, ``Development and application of an aero-hydro-servo-elastic coupling framework for analysis of floating offshore wind turbines,'' \emph{Renewable Energy}, vol. 161, pp. 606--625, 2020.

\bibitem{micallef2017dynamic}
M.~Micallef, T.~Sant, and P.~Mollicone, ``Dynamic analysis of a floating hybrid spar tension leg platform concept for wind monitoring applications in deep sea,'' \emph{IET Renewable Power Generation}, vol.~11, no.~9, pp. 1089--1099, 2017.

\bibitem{liu2024optimization}
T.~Liu, Y.~Liu, S.~Huang, and G.~Xue, ``Optimization of wind-wave hybrid system based on wind-wave coupling model,'' \emph{IET Renewable Power Generation}, 2024.

\bibitem{stansby2024wind}
P.~Stansby and G.~Li, ``A wind semi-sub platform with hinged floats for omnidirectional swell wave energy conversion,'' \emph{Journal of Ocean Engineering and Marine Energy}, vol.~10, no.~2, pp. 433--448, 2024.

\bibitem{imran2017optimal}
R.~M. Imran, D.~M. Akbar~Hussain, M.~Soltani, and R.~M. Rafaq, ``Optimal tuning of multivariable disturbance-observer-based control for flicker mitigation using individual pitch control of wind turbine,'' \emph{IET Renewable Power Generation}, vol.~11, no.~8, pp. 1121--1128, 2017.

\bibitem{RANSLEY201749}
E.~Ransley, D.~Greaves, A.~Raby, D.~Simmonds, M.~M. Jakobsen, and M.~Kramer, ``Rans-vof modelling of the wavestar point absorber,'' \emph{Renewable Energy}, vol. 109, pp. 49--65, 2017.

\bibitem{Robertson2014Definition}
A.~Robertson, J.~Jonkman, M.~Masciola, H.~Song, A.~Goupee, A.~Coulling, and C.~Luan, ``Definition of the semisubmersible floating system for phase ii of oc4,'' National Renewable Energy Lab.(NREL), Golden, CO (United States), Tech. Rep., 2014.

\bibitem{TIAN2023113824}
W.~Tian, Y.~Wang, W.~Shi, C.~Michailides, L.~Wan, and M.~Chen, ``Numerical study of hydrodynamic responses for a combined concept of semisubmersible wind turbine and different layouts of a wave energy converter,'' \emph{Ocean Engineering}, vol. 272, p. 113824, 2023.

\bibitem{Cummins1962The}
W.~E. Cummins, ``The impulse response function and ship motions,'' \emph{Schiffstechnik}, vol.~9, pp. 101--109, 1962.

\bibitem{Prez2009IdentificationOD}
T.~Perez and T.~Fossen, ``Identification of dynamic models of marine structures from frequency-domain data enforcing model structure and parameter constraints,'' \emph{ARC Centre of Excellence for Complex Dynamic Systems and Control}, pp. 1--28, 2009.

\bibitem{da2022dynamics}
L.~Da~Silva, N.~Sergiienko, B.~Cazzolato, and B.~Ding, ``Dynamics of hybrid offshore renewable energy platforms: Heaving point absorbers connected to a semi-submersible floating offshore wind turbine,'' \emph{Renewable Energy}, vol. 199, pp. 1424--1439, 2022.

\bibitem{mishra2021analysis}
M.~K. Mishra \emph{et~al.}, ``Analysis and design of gradient descent based pre-synchronization control for synchronverter,'' \emph{IET Renewable Power Generation}, vol.~15, no.~2, 2021.

\bibitem{li2024highly}
X.~Li, W.~Liu, B.~Liang, Q.~Li, Y.~Zhao, and J.~Hu, ``Highly robust co-estimation of state of charge and state of health using recursive total least squares and unscented kalman filter for lithium-ion battery,'' \emph{IET Renewable Power Generation}, 2024.

\bibitem{HDCB202304001}
Y.~Li, C.~O. Muk, K.~Wang, and L.~Li, ``Effect of the damping parameters of pto systems on the power performance of a new wind-wave energy combined utilization device,'' \emph{Journal of Jiangsu University of Science and Technology (Nature Science Edition)}, vol.~37, pp. 1--9, 2023.

\bibitem{Ghafari2021Numerical}
H.~R. Ghafari, H.~Ghassemi, and G.~He, ``Numerical study of the wavestar wave energy converter with multi-point-absorber around deepcwind semisubmersible floating platform,'' \emph{Ocean Engineering}, vol. 232, p. 109177, 2021.

\bibitem{zhang2020hydrodynamic}
H.~Zhang, B.~Zhou, C.~Vogel, R.~Willden, J.~Zang, and L.~Zhang, ``Hydrodynamic performance of a floating breakwater as an oscillating-buoy type wave energy converter,'' \emph{Applied Energy}, vol. 257, p. 113996, 2020.

\bibitem{anderson2005failures}
B.~D. Anderson, ``Failures of adaptive control theory and their resolution,'' \emph{”Communications in Information and Systems}, vol.~5, no.~1, pp. 1--20, 2005.

\bibitem{larsen2005introduction}
R.~J. Larsen and M.~L. Marx, \emph{An introduction to mathematical statistics}.\hskip 1em plus 0.5em minus 0.4em\relax Prentice Hall Hoboken, NJ, 2005.

\end{thebibliography}

\end{document}